\documentclass[twocolumn,aps,showpacs,preprintnumbers,amsmath,amssymb, superscriptaddress,longbibliography]{revtex4-1}
\pdfoutput=1

\usepackage{bm}
\usepackage{graphicx}
\usepackage{color}

\renewcommand{\vec}{\mathbf}

\begin{document}

\title{Van Hove singularities, chemical pressure and phonons: an angle-resolved photoemission study of KFe$_2$As$_2$ and CsFe$_2$As$_2$}

\author{P. Richard}\email{pierre.richard.qc@gmail.com}
\affiliation{Institut quantique, Universit\'{e} de Sherbrooke, 2500 boulevard de l'Universit\'{e}, Sherbrooke, Qu\'{e}bec J1K 2R1, Canada}
\author{A. van Roekeghem}
\affiliation{CEA, LITEN, 17 Rue des Martyrs, 38054 Grenoble, France}
\author{X. Shi}
\affiliation{Beijing National Laboratory for Condensed Matter Physics, and Institute of Physics, Chinese Academy of Sciences, Beijing 100190, China}
\affiliation{Department of Physics and JILA, University of Colorado and NIST, Boulder, Colorado 80309, USA}
\author{P. Seth}
\affiliation{Centre de Physique Th\'{e}orique, Ecole Polytechnique, CNRS UMR 7644, Universit\'{e} Paris-Saclay, 91128 Palaiseau, France}
\author{B.-Q. Lv}
\affiliation{Beijing National Laboratory for Condensed Matter Physics, and Institute of Physics, Chinese Academy of Sciences, Beijing 100190, China}
\affiliation{School of Physical Sciences, University of Chinese Academy of Sciences, Beijing 100190, China}
\affiliation{Department of Physics, Massachusetts Institute of Technology, Cambridge, MA 02139, USA}
\author{T.~K.~Kim}
\affiliation{Diamond Light Source, Harwell Campus, Didcot, OX11 0DE, United Kingdom}
\author{X.-H. Chen}
\affiliation{Hefei National Laboratory for Physical Sciences at Microscale and Department of Physics, University of Science and Technology of China, Hefei 230026, China}
\affiliation{Collaborative Innovation Center of Advanced Microstructures, Nanjing University, China}
\author{S. Biermann}
\affiliation{Centre de Physique Th\'{e}orique, Ecole Polytechnique, CNRS UMR 7644, Universit\'{e} Paris-Saclay, 91128 Palaiseau, France}
\affiliation{Coll\`{e}ge de France, 11 place Marcelin Berthelot, 75005 Paris, France}
\affiliation{European Theoretical Synchrotron Facility, Europe}
\author{H. Ding}
\affiliation{Beijing National Laboratory for Condensed Matter Physics, and Institute of Physics, Chinese Academy of Sciences, Beijing 100190, China}
\affiliation{School of Physical Sciences, University of Chinese Academy of Sciences, Beijing 100190, China}
\affiliation{Collaborative Innovation Center of Quantum Matter, Beijing, China}

\date{\today}

\begin{abstract}
 We report an angle-resolved photoemission spectroscopy (ARPES) study of KFe$_2$As$_2$ and CsFe$_2$As$_2$, revealing the existence of a van Hove singularity affecting the electronic properties. As a result of chemical pressure, we find a stronger three-dimensionality in KFe$_2$As$_2$ than in CsFe$_2$As$_2$, notably for the 3$d_{z^2}$ states responsible for the small three-dimensional hole-like Fermi surface pocket reported by quantum oscillations. Supported by first-principles calculations, our ARPES study shows that the van Hove singularity previously reported in KFe$_2$As$_2$ moves closer to the Fermi level under negative chemical pressure. This observation, which suggests that the large density-of-states accompanying the van Hove singularity contributes to the large Sommerfeld coefficient reported for the AFe$_2$As$_2$ (A = K, Rb, Cs) series, is also consistent with the evolution of the inelastic scattering revealed by transport under external pressure, thus offering a possible interpretation for the origin of the apparent change in the superconducting order parameter under pressure. We find that the coherent spectral weight decreases exponentially upon increasing temperature with a characteristic temperature $T^*$. We speculate how the low-energy location of the van Hove singularity and the presence of a low-energy peak in the phonon density-of-states can relate to the high-temperature crossover observed in various electronic and thermodynamic quantities.  
\end{abstract}



\maketitle

\section{Introduction}

Conventional metals, in which delocalized electrons are associated with Bloch wave functions, are well described by the Fermi liquid theory. Departure from this behavior, usually attributed to electronic correlations, include the renormalization of electronic bands and the loss of spectral coherence, with signatures on electronic transport and thermodynamic properties. Among other Fe-based superconductors, AFe$_2$As$_2$ (A = K, Rb, Cs) is widely believed to be more correlated than most of its cousins. This is based partly on theoretical grounds indicating that the strength of the electronic correlations should increase towards half filling of the $d$ shell \cite{WernerPRL101, HauleNJP11, WernerNatPhys8, deMedici_PRL107, vanRoekeghemCR17}, and on experimental results suggesting heavy masses and a crossover from coherent to incoherent occurring as we increase temperature across the crossover temperature $T^*$ \cite{Hardy_PRL111, YP_Wu_PRL116, Wiecki_PRB97}. 

Whether the electronic correlations are responsible for the superconducting properties of these materials is debated. Although there is no consensus on the details of the gap structure of AFe$_2$As$_2$ (A = K, Rb, Cs), several experimental reports suggest the presence of nodes, notably from angle-resolved photoemission spectroscopy (ARPES) \cite{Okazaki_Science337}, thermal conductivity \cite{JK_DongPRL2010,ReidPRL109,XC_HongPRB87,AF_WangPRB89, Z_Zhang_PRB91}, specific heat \cite{Abdel-HafiezPRB85,AF_WangPRB87,KittakaJPSJ2014} and penetration depth \cite{Hashimoto_PRB82,Mizukami_PRB89}. In a series of papers, Tafti \textit{et al.} reported a sudden reversal in the pressure dependence of the superconducting temperature $T_c$ in AFe$_2$As$_2$ \cite{Tafti_nphys9,Tafti_PRB89,Tafti_PRB91}, which was interpreted in terms of a change in the order parameter from $d$-wave to $s$-wave at a critical pressure $P_c$. Similar anomaly was also reported in NMR studies under high-pressure \cite{PS_WangPRB93,Wiecki_PRB97}. The pressure studies of Tafti \textit{et al.} also showed that while the inelastic resistivity, defined as the resistivity $\rho(T)$ minus the residual resistivity $\rho_0$, varies linearly with pressure above $P_c$, a clear rise in $\rho(T)-\rho_0$ was found below $P_c$, indicating the appearance of an additional inelastic scattering channel at low pressure and ambiant pressure conditions \cite{Tafti_PRB89,Tafti_PRB91}. The origin of the additional inelastic component at low pressure, of its pressure dependence and of its link to nodal superconductivity are unclear. 

Due to its ability to determine the momentum-resolved electronic structure not only at the Fermi energy ($E_F$) but also in a wide energy range below $E_F$, ARPES is a powerful tool for addressing the origin of inelastic scattering in AFe$_2$As$_2$ and determine the possible implication of strong electronic correlations. Here we show that while the presence of strong electronic correlations in AFe$_2$As$_2$ is undeniable, other parameters such as a van Hove singularity in the low-energy electronic structure, play critical roles in shaping the rather peculiar properties of these materials. 

The main experimental results of this paper are presented in Sections~\ref{Cs_section} and \ref{temperature_section}. In Section~\ref{Cs_section}, we compare ARPES data recorded on KFe$_2$As$_2$ and CsFe$_2$As$_2$, with the chemical substitution of K by Cs acting similarly to the application of a negative pressure \cite{Tafti_PRB89,Tafti_PRB91}. We show that the 3$d_{z^2}$ states responsible for the small three-dimensional hole pocket reported in quantum oscillation measurements \cite{JPSConf2014_Zocco} are less $k_z$-dispersive in CsFe$_2$As$_2$ than in KFe$_2$As$_2$, but that they are located closer to $E_F$. Interestingly, the van Hove singularity located midway between the $\Gamma$ and M points, directly on the nodal line of the $s_{\pm}$ function, and identified previously as candidate for the nodal behavior in KFe$_2$As$_2$ \cite{DL_Fang_vHs}, also moves closer to $E_F$ with decreasing chemical pressure. Our observation offers a natural explanation 1)~for the lower Sommerfeld coefficient in KFe$_2$As$_2$ as compared to CsFe$_2$As$_2$ \cite{Abdel-HafiezPRB85,AF_WangPRB87,Hardy_PRB94,Eilers_PRL116}, 2)~for the decrease of the non-linear component of the inelastic scattering with decreasing pressure towards $P_c$, and 3)~for the change in the superconducting order parameter at $P_c$. In Section~\ref{temperature_section}, we present the temperature evolution of the ARPES spectrum of KFe$_2$As$_2$ up to 250~K. We show that the electronic spectral weight decreases exponentially with a characteristic temperature consistent with $T^*$ reported with other techniques \cite{YP_Wu_PRL116, Wiecki_PRB97}. We discuss alternatives to the widely accepted strong correlation origin of $T^*$ in terms of the the phonon density-of-states (PDOS) and of the complicated low-energy electronic structure from both sides of $E_F$ revealed by the ARPES data. 

\section{experiment}

High-quality single crystals of KFe$_2$As$_2$ and CsFe$_2$As$_2$ were grown by conventional self-flux methods \cite{AF_WangPRB89}. ARPES measurements were performed at Beamline I05 of Diamond Light Source equipped with a VG-Scienta R4000 analyzer. The energy and angular resolutions for the angle-resolved data were set at 8 - 30~meV and 0.2$^{\textrm{o}}$, respectively. The samples were cleaved \emph{in situ} and measured at 7~K in a vacuum better than $5\times 10^{-11}$ Torr. Additional measurements were performed on KFe$_2$As$_2$ samples cleaved at either 10~K or 240~K and measured between 20~K and 250~K, using $\sigma$-polarized photons of 70.5~eV energy, which probe the plane corresponding to a perpendicular momentum $k_z=0$. Throughout the paper, we label the momentum values with respect to the 1~Fe/unit-cell Brillouin zone (BZ), and use $c^{\prime} = c/2$ as the distance between two Fe planes.

\section{K$\textrm{Fe}_2$$\textrm{As}_2$ vs $\textrm{CsFe}_2$$\textrm{As}_2$: tuning the van Hove singularity with chemical pressure\label{Cs_section}}

\begin{figure}[!t]
\begin{center}
\includegraphics[width=\columnwidth]{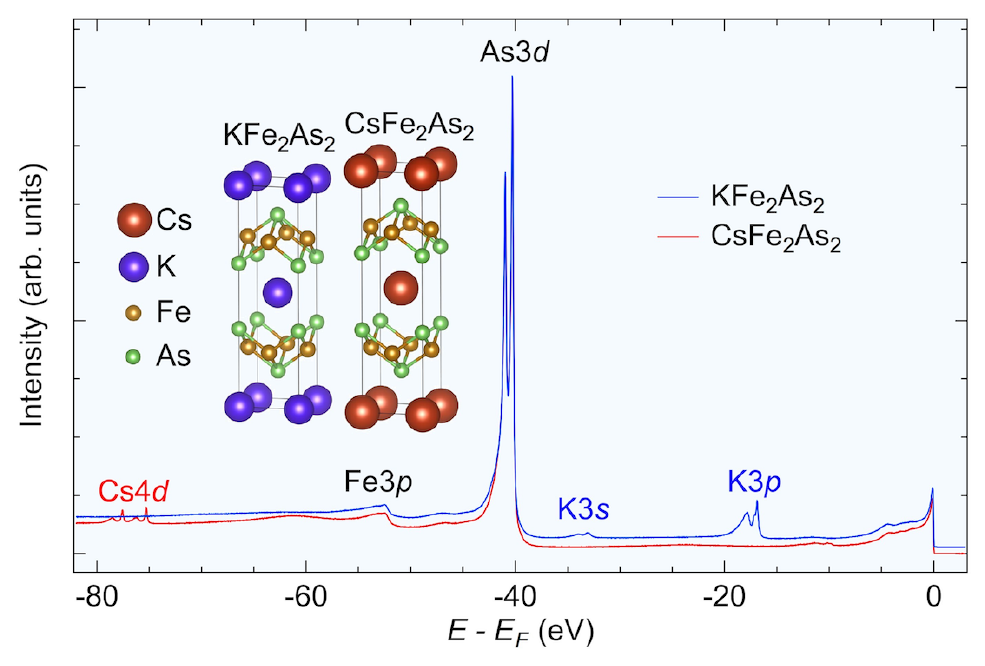}
\end{center}
\caption{\label{Core}(Color online) Core level spectra of KFe$_2$As$_2$ and CsFe$_2$As$_2$. The corresponding crystal structures are shown in inset.}
\end{figure}

Photoemission is an experimental technique in which electrons are excited out of a surface exposed to a monochromatic light beam. Using the law of conservation of energy, photoemission allows us to trace an electronic spectrum characteristic of each material up to $E_F$. In Fig. \ref{Core} we compare the core level spectra of KFe$_2$As$_2$ and CsFe$_2$As$_2$. Without any surprise, the spectrum of KFe$_2$As$_2$ contains peaks unique to K, notably the K$3s$ and K$3p$ states, while the spectrum of CsFe$_2$As$_2$ exhibits Cs$4d$ states not observed in KFe$_2$As$_2$. The rest of the spectra are very similar, and we have to go into the details to see some differences.

\begin{figure}[!t]
\begin{center}
\includegraphics[width=\columnwidth]{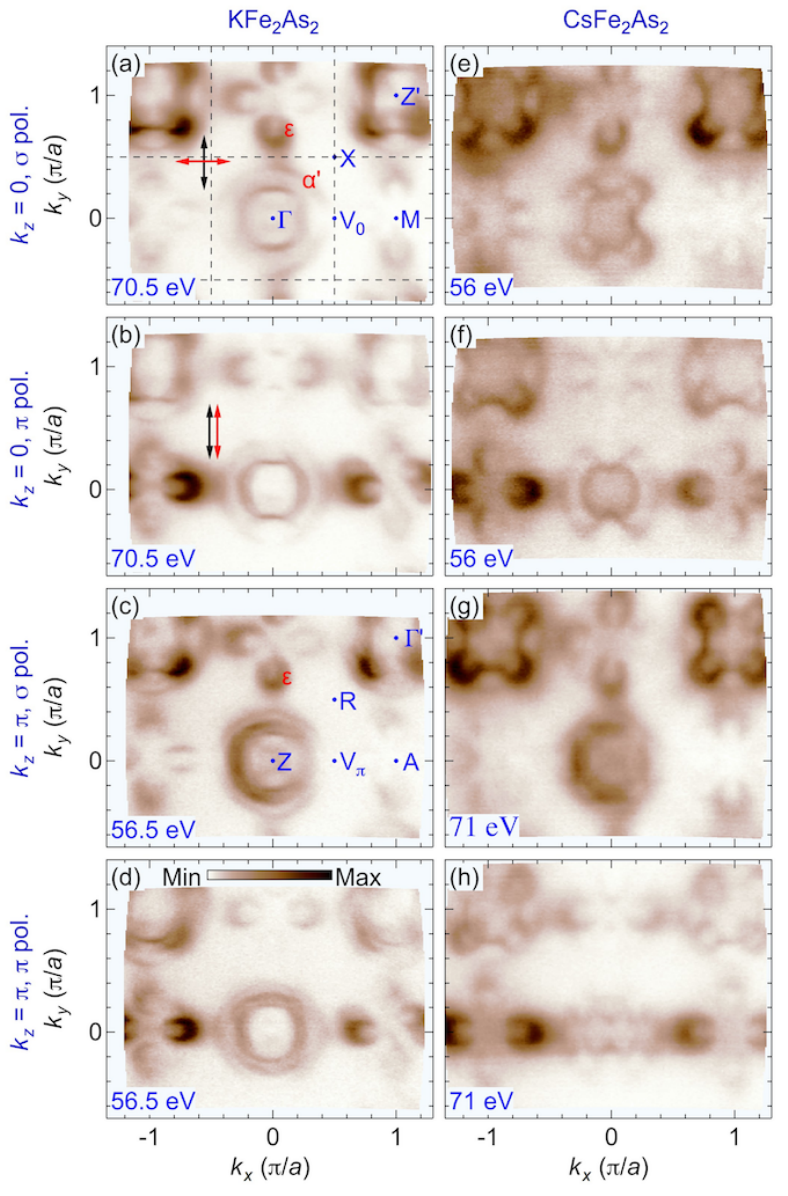}
\end{center}
\caption{\label{FSs}(Color online) Comparison of the FSs ($\pm 5$ meV integration) of KFe$_2$As$_2$ (left column) and CsFe$_2$As$_2$ (right column) at different $k_z$ values and under both $\sigma$ and $\pi$ polarizations. While the black double-arrow indicates the orientation of the analyzer slit, the red double-arrows in (a) and (b) define the orientation of the polarization for the $\sigma$ and $\pi$ configurations, respectively. The locations of high-symmetry points and of the $\varepsilon$ and $\alpha^{\prime}$ FSs are shown in (a) for the $k_z=0$ plane and in (c) for the $k_z=\pi$ plane. The dashed lines in (a) correspond to the zero gap lines of the $s_{\pm}$ gap function. The color code for the photoemission intensity is given in panel (d).}
\end{figure}

ARPES is an advanced photoemission technique in which not only the energy but also the momentum of the photoemitted electrons are measured. Taking also into account the law of conservation of momentum, ARPES probes directly the momentum-resolved electronic band structure, including constant energy maps. We show in Fig. \ref{FSs} a series of Fermi surface (FS) mappings recorded on KFe$_2$As$_2$ and CsFe$_2$As$_2$ in the $k_z=0$ and $k_z=\pi$ planes, under both $\sigma$ and $\pi$ configurations of polarization. Although the FSs of KFe$_2$As$_2$ and CsFe$_2$As$_2$ have already been reported and described in previous ARPES  \cite{Sato_PRL2009,Yoshida_JCPS72,Yoshida_FP2,Okazaki_Science337,Kong_PRB92} and quantum oscillation studies \cite{Terashima_JPSJ79,Terashima_PRB87,JPSConf2014_Zocco,Terashima_PRB89}, some comments need to be made. The FS patterns recorded around the M $(1,0,k_z=0)$ and A $(1,0,k_z=\pi)$ points are practically equivalent, indicating that there is not much $k_z$ dispersion along M-A. The situation is different along the $(0,0,k_z)$ direction. As expected, the intensity patterns obtained around $(0,0,k_z)$ are different at $k_z=0$ ($\Gamma$) and $k_z=\pi$ ($Z$). For KFe$_2$As$_2$ at $k_z=0$ and with $\sigma$ polarization (Fig. \ref{FSs}(a)), we distinguish clearly two hole bands. The inner one is nearly doubly-degenerate, but $\pi$ polarization allows to resolve the two components (Fig. \ref{FSs}(b)). These three hole FS pockets are better resolved at $k_z=\pi$ (Figs. \ref{FSs}(c) and \ref{FSs}(d)). An additional but small FS is detected clearly  at $(0,0,\pi)$. As we explain below, this extra FS pocket is also present at $k_z=0$ and was previously attributed to a surface state \cite{Yoshida_JCPS72,Yoshida_FP2}. 

Unexpectedly, the experimental FS intensity patterns around $Z=(0,0,\pi)$ and $Z'=(\pi,\pi,0)$, which should be equivalent considering the body-centered structure of KFe$_2$As$_2$, are clearly different. Similarly, the FS intensity patterns around $\Gamma=(0,0,0)$ and $\Gamma'=(\pi,\pi,\pi)$ differ. The pattern found experimentally around $Z'$ is more consistent with the one predicted theoretically by the local density approximation (LDA) combined with dynamical mean field theory (DMFT) for the $\Gamma$ point, with one of the FS pocket exhibiting a rather star-like shape (see Fig. 2(b)\footnote{We caution that the axes in Ref. \cite{Backes_NJP16} are rotated by 45$^{\circ}$ with respect to ours} of Ref. \cite{Backes_NJP16}). Moreover, a large FS pocket at $(\pi,\pi,k_z)$, with the same size as the large one observed at $(0,0,k_z)$, suggests a weak band folding from $(0,0,k_z)$ to $(\pi,\pi,k_z)$, which could possibly arise from a surface contribution or a $k_z$ averaging effect. It is also worth mentioning that some of the FS intensity patterns obtained by ARPES are very similar to the peculiar ones calculated by LDA+DMFT and displayed in Fig. 6 of Ref. \cite{Backes_NJP16}, notably for the $d_{xz}/d_{yz}$ bands.

\begin{figure*}[!t]
\begin{center}
\includegraphics[width=\textwidth]{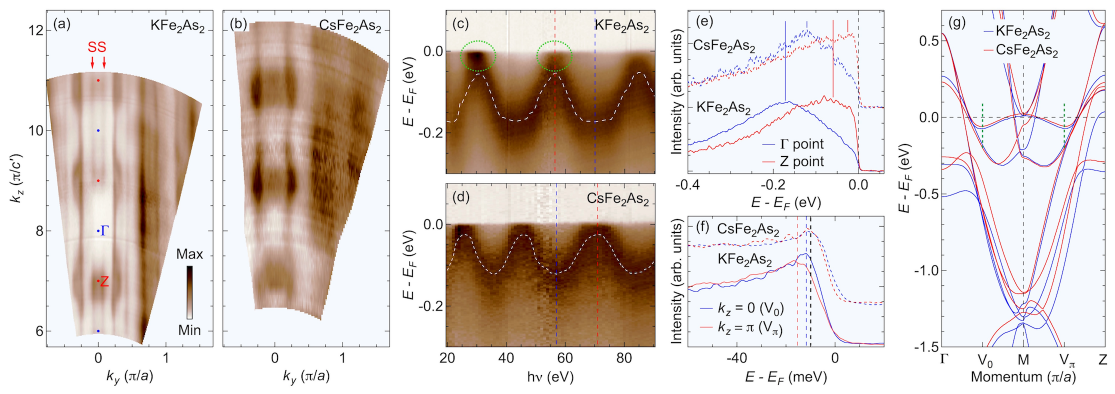}
\end{center}
\caption{\label{kzmap}(Color online) (a), (b) FS intensity map ($\pm 5$ meV integration) in the $k_y-k_z$ for KFe$_2$As$_2$ and CsFe$_2$As$_2$, respectively. The photoemission intensity color code is given in panel (a), as well as the $k_z$ values corresponding to the $\Gamma$ (blue) and Z (red) points. The data have been recorded over the same 20-90 eV photon energy range ($\sigma$ polarization). (c), (d) Photon energy dependence of the normal incidence EDC in KFe$_2$As$_2$ and CsFe$_2$As$_2$, respectively. The dashed white curves represent the dispersion of the $d_{z^2}$ band extracted from the ARPES data. The vertical blue and red dashed lines indicate the photon energies, corresponding respectively to $k_z=0$ and $k_z=\pi$, at which the EDCs in panel (e) have been recorded. The green circles are used to emphasize a surface sate. (e) Comparison between the EDCs of KFe$_2$As$_2$ and CsFe$_2$As$_2$ measured at the $\Gamma$ (blue) and Z (red) points. The vertical lines mark the peak positions. (f) Comparison between the EDCs of KFe$_2$As$_2$ and CsFe$_2$As$_2$ measured at V$_0$ (blue) and V$_{\pi}$ (red) points. The vertical dashed lines in red and blue mark the peak positions in KFe$_2$As$_2$ whereas the black dashed line indicates the peak position in CsFe$_2$As$_2$. (g) LDA bands of KFe$_2$As$_2$ and CsFe$_2$As$_2$ calculated along $\Gamma$-M-$Z$ at $k_z=0$.} 
\end{figure*}


In order to obtain a more detailed representation of the dispersion along the $k_z$ direction, we recorded the photon energy ($h\nu$) dependence of a cut along $\Gamma$-M. The results for $h\nu$ varying between 20 and 90 eV are displayed in Figs. \ref{kzmap}(a) and \ref{kzmap}(b) for KFe$_2$As$_2$ and CsFe$_2$As$_2$, respectively. Within the sudden approximation and the so-called three-step model with free-electron final state for the photoemission process \cite{RichardJPCM27review}, the momentum $k_z$ can be approximated by $|\vec{k}_z|=(\sqrt{2m}/\hbar)[E_{kin}+U_0-(\hbar\vec{k_{||}})^2/{2m}]^{1/2}$, where $E_{kin}$ is the kinetic energy of the photoemitted electrons, $\vec{k_{||}}$ the in-plane component of the electron before the photoemission, $m$ the free-electron mass and $U_0$ the inner potential used as free parameter. The conversion from $h\nu$ to $k_z$ works well when choosing $U_0=12$ eV for both KFe$_2$As$_2$ and CsFe$_2$As$_2$, although the same $h\nu$ values correspond to different $k_z$ due to the difference in the $c$ axis parameter. 

Most of the spectral features do not disperse significantly along $k_z$, except two. The first $k_z$-dispersive feature is the $\alpha$ band, which is mainly formed by the even combination of the $d_{xz}$ and $d_{yz}$ orbitals, with a sizable $d_{z^2}$ component. In particular, this band gives rise to a larger FS around the Z point than around $\Gamma$, as shown in Figs. \ref{kzmap}(a) and \ref{kzmap}(b). Stronger photoemission intensity is also found around the Z point, for $k_y =0$. Interestingly, the size of this spot of intensity (the second $k_z$-dispersive feature) is comparable to that of the non-$k_z$-dispersive SS band marked by arrows in Fig. \ref{kzmap}(a), which has been attributed previously to a surface state \cite{Yoshida_JCPS72}. To gain more insight about this issue, we show in Figs. \ref{kzmap}(c) and \ref{kzmap}(d) the $h\nu$ dependence of the normal photoemission energy distribution curve (EDC) in KFe$_2$As$_2$ and CsFe$_2$As$_2$, respectively. We observe a strong $k_z$-dispersive feature attributed to the $d_{z^2}$ band. As emphasized by green circles in Figs. \ref{kzmap}(c), the surface state becomes stronger when the $d_{z^2}$ band approaches $E_F$, suggesting that the surface probably originates from the $d_{z^2}$ orbital on the top layer surface.

As we discussed above, the ARPES measurements possibly suffer from a $k_z$ averaging effect. Although this would not affect much the weakly $k_z$-dispersive bands, the effect should be sizable for the $d_{z^2}$ band. In particular, we suspect that such $k_z$ averaging effect should reduce the apparent dispersion along $k_z$, and we cannot exclude the possibility that this band crosses $E_F$ at the Z point, forming a three-dimensional FS pocket, as proposed by quantum oscillation experiments \cite{JPSConf2014_Zocco}. In any case, the $d_{z^2}$ band is certainly affected by the chemical pressure, as illustrated by the EDCs at the $\Gamma$ and Z points shown in Fig. \ref{kzmap}(e). Indeed, the top of the band, as determined by ARPES, shifts from 56 meV below $E_F$ in KFe$_2$As$_2$ to 25 meV below $E_F$ in CsFe$_2$As$_2$. Moreover, as expected from the smaller inter-layer distance in KFe$_2$As$_2$, the $k_z$ variation for the $d_{z^2}$ band is larger for this material, for which we record a 116 meV difference between the peak positions at $\Gamma$ and Z, as compared to 97 meV in CsFe$_2$As$_2$. This observation is consistent with our LDA calculations, which show stronger three-dimensionality in KFe$_2$As$_2$ than in CsFe$_2$As$_2$.

We note that the presence of a $d_{z^2}$ band is a factor that may contribute to reduce the strength of the electronic correlations. This situation is similar to that of BaCr$_2$As$_2$, which is symmetrical to the BaFe$_2$As$_2$ with respect to the half-filled $3d$, but which is much less correlated than its ferropnictide cousin \cite{RichardPRB95, Nayak_PNAS114}. It was argued that the admixture of the $t_{2g}$ ($d_{xy}$, $d_{xz}$ and $d_{yz}$) and $e_g$ ($d_{z^2}$ and $d_{{x^2}-{y^2}}$) orbitals found in BaCr$_2$As$_2$ \cite{RichardPRB95} was detrimental to high-temperature superconductivity \cite{JP_HuPRX5} and that the increase $p-d$ hybridization, which should be affected by an increased $d_{z^2}$ character, can affect the electronic correlations by modifying the effective electronic count in the $3d$ shell \cite{Razzoli_PRB91}. Nevertheless, a local density approximation combined with dynamical mean field theory study rather suggests a slight increase in the electronic correlations by going along the K, Rb, Cs series, mainly attributed to more localized $d_{xy}$ orbitals resulting from a larger $a$ lattice parameter \cite{Backes_PRB92}.
 
A previous ARPES and scanning tunneling microscopy (STM) study revealed the existence of a van Hove singularity in KFe$_2$As$_2$ located only a few meV below $E_F$ \cite{DL_Fang_vHs}. This van Hove singularity was also observed in first-principles calculations, though at higher energy \cite{Drechsler_PSSB254}. The momentum location of the van Hove singularity in the $k_z=0$ plane (V$_0$)  and $k_z=\pi$ plane (V$_{\pi}$) are indicated in Figs. \ref{FSs}(a) and \ref{FSs}(c), respectively. In Fig. \ref{kzmap}(f) we compare the EDCs measured at both V$_0$ and V$_{\pi}$ in KFe$_2$As$_2$ and CsFe$_2$As$_2$. The van Hove singularity in KFe$_2$As$_2$ moves from -15 meV at V$_{\pi}$ to -12 meV at V$_0$. As expected, the dispersion of the van Hove singularity along $k_z$ is even smaller in CsFe$_2$As$_2$, and we find -10 meV for its energy position at both V$_0$ and V$_{\pi}$, which is closer to $E_F$ than in KFe$_2$As. These results are consistent with the -15 meV and -11 meV energies reported in Ref. \cite{Drechsler_JSNM31} for KFe$_2$As$_2$ and Cs$_{0.94}$K$_{0.06}$Fe$_2$As$_2$. Our result is also qualitatively consistent with the LDA calculations displayed in Fig. \ref{kzmap}(g), which show that the van Hove singularity is closer to $E_F$ in CsFe$_2$As$_2$ than in KFe$_2$As$_2$ by 15 meV at both V$_0$ and V$_{\pi}$. Obviously, the location of the van Hove singularity is critical for estimating the contribution of the density-of-states at the Fermi level entering the computation of the Sommerfeld coefficient, and a lack of knowledge of its location can lead to overestimated electronic correlation strengths \cite{Drechsler_PSSB254}. Although the Fe-based superconductors exhibit relatively strong electronic correlations \cite{vanRoekeghemCR17}, the large density-of-states due to the proximity of the van Hove singularity likely contributes to 
the large Sommerfeld coefficients reported for KFe$_2$As$_2$ and CsFe$_2$As$_2$, rather than strong electronic correlations alone. Such a conclusion was also drawn for the isostructural compound TlNi$_2$Se$_2$ \cite{Nan_XuPRB92}. Heavy fermion behavior in TlNi$_2$Se$_2$ was inferred from specific heat measurements \cite{H_WangPRL111}, but ARPES and LDA calculations later revealed a weakly correlated electronic structure containing a van Hove singularity near $E_F$ \cite{Nan_XuPRB92}. Our current results suggest an even larger Sommerfeld coefficient in CsFe$_2$As$_2$ than in KFe$_2$As$_2$, in agreement with specific heat studies \cite{Abdel-HafiezPRB85,AF_WangPRB87,Hardy_PRB94} reporting Sommerfeld coefficients of 94 mJ/molK$^{2}$ in KFe$_2$As$_2$ \cite{Abdel-HafiezPRB85} and 184 mJ/molK$^{2}$ in CsFe$_2$As$_2$ \cite{AF_WangPRB87}. Since the van Hove singularity is located not too far from $E_F$, it should be regarded as an important source of inelastic scattering that should fade as it moves away from $E_F$ with chemical pressure (or external pressure), in agreement with the transport measurements of Tafti \textit{et al.} \cite{Tafti_PRB89,Tafti_PRB91}. 

We now address the role of the van Hove singularity in shaping the order parameter. We first recall that the sudden reversal of $T_c$ under pressure, attributed to a change in the order parameter, coincides with the loss of the non-linear component in the inelastic resistivity \cite{Tafti_nphys9,Tafti_PRB89,Tafti_PRB91}, which we here relate to a van Hove singularity. As noticed in a previous STM and ARPES study, the van Hove singularity is located directly on the nodal line of the $s_{\pm}$ gap function \cite{DL_Fang_vHs}, not very far from the tip of the M-off-centered, petal-shape $\varepsilon$ hole pocket \cite{Sato_PRL2009}. Interestingly, the tail of the van Hove singularity peak extends to zero bias in the STM data, leading to a non-zero bias even in the superconducting state \cite{DL_Fang_vHs}. It was argued that the proximity of the tip of the $\varepsilon$ pocket to the van Hove singularity may impose a zero amplitude to the gap at the tip position \cite{DL_Fang_vHs}, such as observed in heavily-hole-doped Ba$_{0.1}$K$_{0.9}$Fe$_2$As$_2$ \cite{Nan_XuPRB88}. In this scenario, the node in the superconducting gap would appear as symmetry-imposed as long as the non-linear inelastic resistivity component associated with the van Hove singularity prevails. When this component becomes negligible at $P_c$ as a result of a too large shift of the van Hove singularity away from $E_F$, the node in the gap structure is lifted, thus explaining the sudden change in the pressure dependence of $T_c$ \cite{Tafti_nphys9,Tafti_PRB89,Tafti_PRB91}, a change that does not require any modification in the topology of the Fermi surface, in agreement with the smooth evolution of the Hall coefficient with pressure \cite{Tafti_nphys9,Tafti_PRB89,Tafti_PRB91}, as well as with quantum oscillation measurements on KFe$_2$As$_2$ under pressure \cite{Terashima_PRB89}. 

\section{Effect of thermal broadening on the electronic properties\label{temperature_section}}

\begin{figure}[!t]
\begin{center}
\includegraphics[width=\columnwidth]{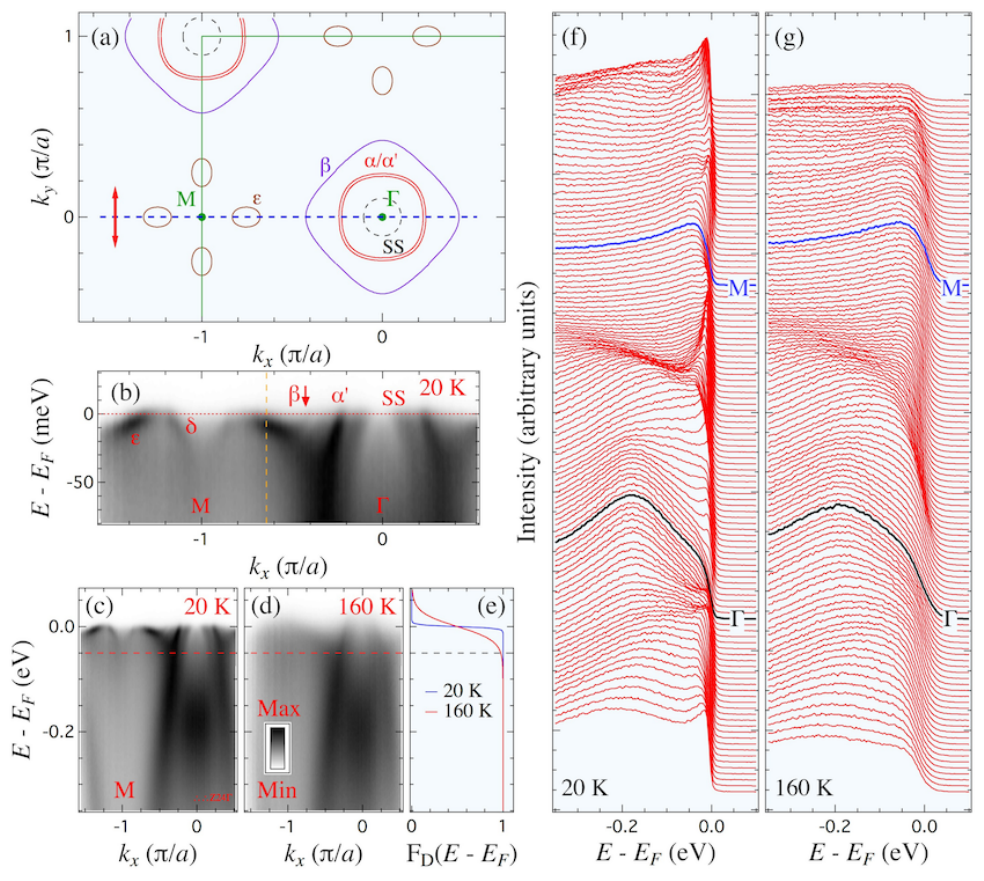}
\end{center}
\caption{\label{FD}(Color online) (a) Schematic FS of KFe$_2$As$_2$. The dashed circle labeled SS represents a surface state \cite{Yoshida_JCPS72} discussed in Section \ref{Cs_section}. The green line is the 1 Fe/unit cell BZ boundary. The red double-arrow represents the direction of the light polarization while the blue dashed line indicates the cut along which all data have been recorded. (b) Low-energy zoom of the ARPES intensity plot along $\Gamma$-M. The color scale is the same as in panel (d). The arrow indicates the weak $\beta$ band with $d_{xy}$ orbital character. The dotted horizontal line refers to $E_F$. While the Greek symbols refer to bulk electronic bands, SS is associated to a surface state. The vertical dashed line indicates the momentum location of the EDCs in Fig. \ref{Coherence}. (c) and (d) ARPES intensity plots recorded at 20 K and 160 K, respectively, along the cut indicated in (a). The color code for the intensity is given in panel (d). The dashed lines indicate 50 meV below $E_F$. (e) Fermi-Dirac function (F$_{\textrm{D}}$) at 20 K and 160 K. The dashed line located 50 meV below $E_F$ is a guide to indicate the energy range for which the ARPES data are directly affected by the F$_{\textrm{D}}$ cutoff. (f) EDC plot corresponding to the data in panel (c), recorded at 20 K. (g) Same as (f) but for the data in panel (d), recorded at 160 K.}
\end{figure}

The schematic FS of KFe$_2$As$_2$ at $k_z=0$, consistent with early studies of this material \cite{Sato_PRL2009,Yoshida_JCPS72}, is illustrated in Fig. \ref{FD}(a), along with the experimental configuration for the temperature-dependent measurements. The data presented in this section have been recorded along the dashed line in Fig.~\ref{FD}(a), corresponding to the $\Gamma$-M direction. In Fig.~\ref{FD}(b) we show a low-energy zoom of the ARPES cut measured at 20~K and identify the low-energy features. The inner hole band observed near $\Gamma$ is the surface state with $d_{z^2}$ character that we discussed in the previous section. The strongest photoemission intensity is found for the $\alpha^{\prime}$ band, which is the odd combination of the nearly-degenerate $d_{xz}/d_{yz}$ bands. Although its intensity should be maximized in the current experimental configuration ($\sigma$ polarization), the $\beta$ band with $d_{xy}$ orbital character, which is marked by an arrow in Fig.~\ref{FD}(b), is barely visible. Also weak in intensity is the $\delta$ band, which also carries a dominant $d_{xy}$ character and forms the section of the holelike $\varepsilon$ FS pocket closest to the M point. Finally, we see very clearly the $\varepsilon$ band forming the outer part of the $\varepsilon$ FS pocket (towards $\Gamma$), which has a main $d_{xz}$ character along the $\Gamma$-M direction.

In Figs. \ref{FD}(c) and \ref{FD}(d) we compare the ARPES intensity plots over a slightly wider energy range, as measured at 20~K and 160~K, respectively. The corresponding EDC curves are displayed in Figs. \ref{FD}(f) and \ref{FD}(g), respectively. All spectral features broaden with increasing temperature, which is easy to understand in terms of an imaginary part of the self-energy $\Sigma$ that increases with temperature. Interestingly, a sharp contrast can be seen between the $\alpha^{\prime}$ band and the bands located near the M point. While the former band remains distinguishable at 160~K, the spectral intensity of the bands around the M point is strongly suppressed. This effect is partly due to the Fermi-Dirac cutoff. In Fig. \ref{FD}(d) we compare the Fermi-Dirac function (F$_{\textrm{D}}$) at 20~K and 160~K. At 160~K, the F$_{\textrm{D}}$ function deviates appreciably from 1 around 50~meV below $E_F$. Since the bands forming the $\varepsilon$ pocket at M are very shallow, their spectral intensity is naturally more affected and they seem washed out, as we will demonstrate below.  

\begin{figure}[!t]
\begin{center}
\includegraphics[width=\columnwidth]{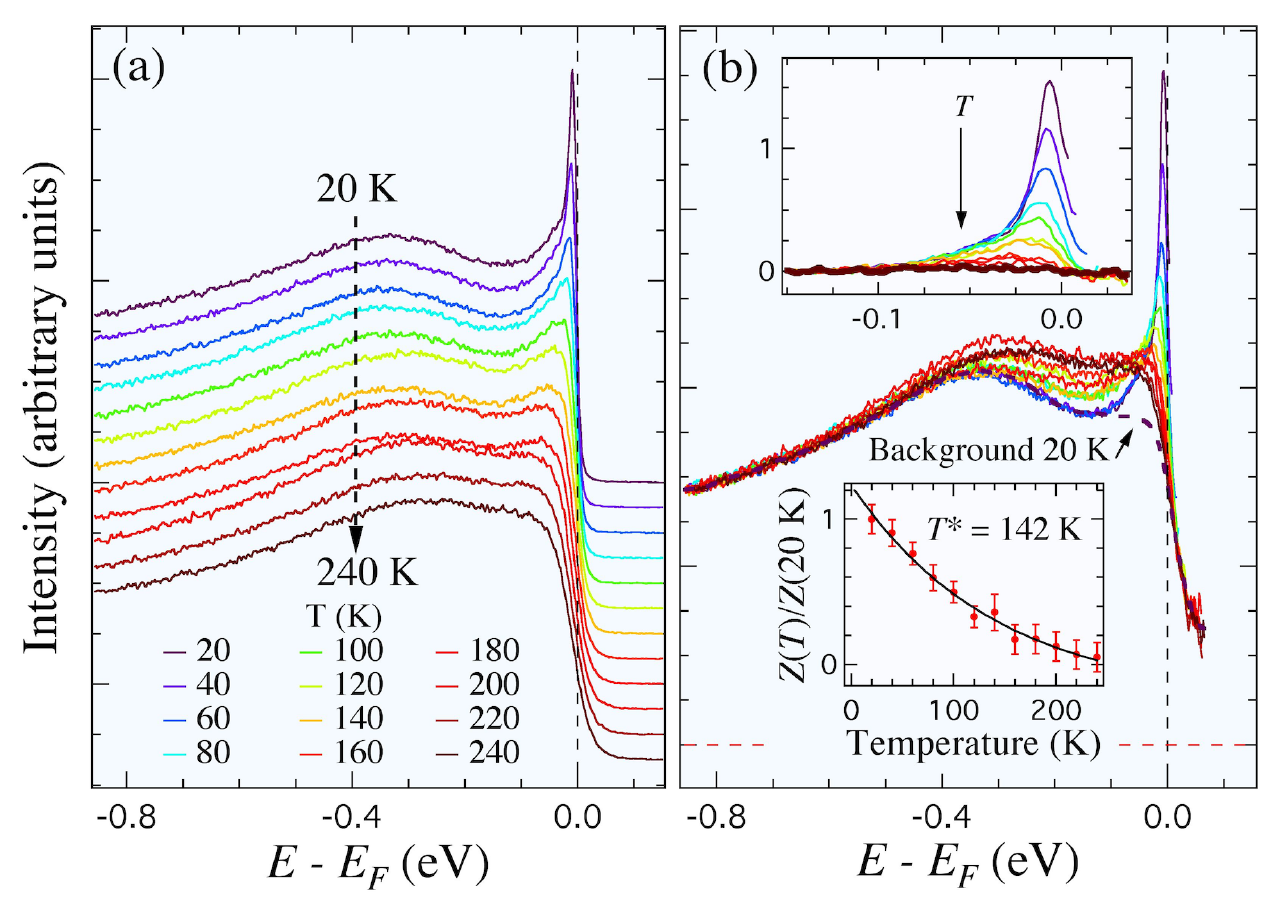}
\end{center}
\caption{\label{Coherence}(Color online) (a) Temperature evolution of the EDCs recorded at the momentum location indicated by the orange vertical dashed line in Fig. \ref{FD}(b). The EDCs have been shifted for a better visualization. (b) Same as in (a) but after division of the EDCs by the F$_{\textrm{D}}$ function convoluted by the instrumental resolution function. The dashed curve indicates the background at 20 K from which the EDC at the same temperature was subtracted in order to extract the coherent component. The top inset corresponds to the coherent component obtained by subtracting the EDCs by a background (see the text). The bottom inset is the coherent factor normalized at 20 K corresponding to the area below the curves in the top inset. The black curve is an exponential fit of the data. }
\end{figure}

\begin{figure*}[!t]
\begin{center}
\includegraphics[width=\textwidth]{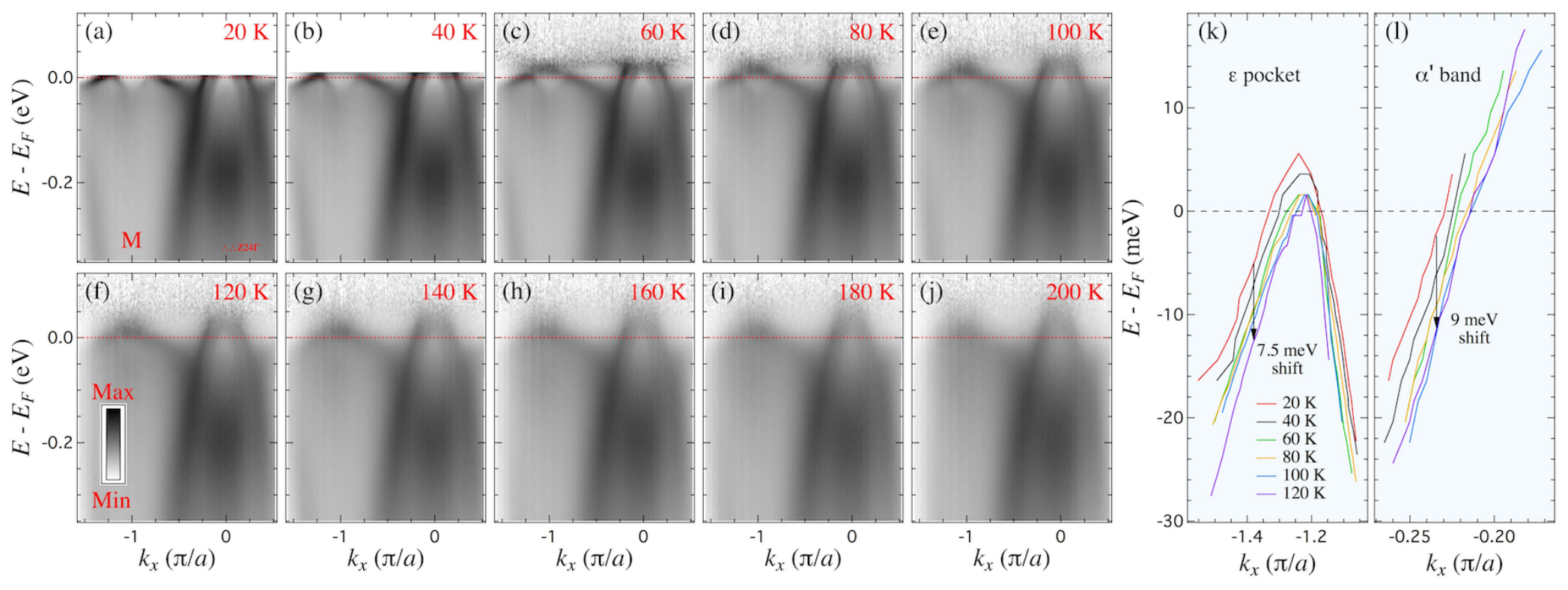}
\end{center}
\caption{\label{Tcuts}(Color online) (a)-(j) ARPES intensity plots of KFe$_2$As$_2$ along the $\Gamma$-M direction, recorded at different temperatures. The color code is given in panel (f). All spectrum have been divided by the F$_{\textrm{D}}$ function convoluted with the instrumental resolution function. (k) Extracted band dispersion near $E_F$ for the bands forming the $\varepsilon$ pocket, recorded at different temperatures. (l) Same as (k) but for the $\alpha$ band.} 
\end{figure*}

In addition to becoming broader, the spectral features become fainter with increasing temperature. To illustrate this effect, we focus on the $\varepsilon$ band, which is well isolated and thus more suitable for a quantitative analysis. We show in Fig. \ref{Coherence}(a) the temperature evolution of the EDC on the $\varepsilon$ band, at the momentum position indicated by a vertical dashed line in Fig. \ref{FD}(b). The EDC at 20 K shows a very sharp and intense peak at 8 meV of binding energy. The intensity of this peak decreases progressively with increasing temperature, and it is barely distinguishable above 180 K. In order to minimize complications related to the proximity of that peak to $E_F$, we first divide all the ARPES spectra by the Fermi-Dirac function convoluted by the instrumental resolution function. The results are displayed in Fig. \ref{Coherence}(b). The conclusion regarding the decrease of the peak intensity with temperature is the same. Interestingly, incoherent spectral intensity builds up between the $\varepsilon$ peak and the broad spectral feature at 330 meV corresponding to the high-energy tail of the $\alpha^{\prime}$ band. Obviously, the $\varepsilon$ band has gone from a highly coherent state at low temperature to a more incoherent one at high temperature. This conclusion is consistent with recent Knight shift and resistivity measurements suggesting a coherent-incoherent crossover at a characteristic temperature $T^*$ estimated to $165\pm 25$ K in Ref. \cite{YP_Wu_PRL116} and $145\pm 20$ K in Ref. \cite{Wiecki_PRB97}.
We note however that the crossovers observed $165$ K in Ref. \cite{YP_Wu_PRL116} and at $145\pm 20$ K in Ref. \cite{Wiecki_PRB97} correspond rather to crossovers between two linear-in-temperature regimes of the resistivity. Coherence in the sense of Fermi liquid $T^2$ behavior is in fact lost at much lower temperatures, of the order of a few tens of K %
\footnote{See e.g. Fig.~3 of Ref.~\cite{TaufourPRB89}, where deviations from $T^2$ behavior of the resistivity appear at temperatures of the order of 60-70 K.}.

For a more quantitative description of the crossovers, we extracted the coherent peak at each temperature after removing a background such as the one illustrated in Fig. \ref{Coherence}(b) for the 20 K spectrum. The results, displayed in the top inset of Fig. \ref{Coherence}(b), clearly show a loss of
spectral weight as temperature increases. The coherent weight $Z(T)$ at each temperature corresponds to the area below the subtracted curves. As illustrated in the bottom inset of Fig. \ref{Coherence}(b), the coherent weight decreases exponentially with temperature, with a characteristic temperature $T^*=142$ K obtained by fitting the data. Since the choice of background used for extracting the coherent components is far from unique, we estimate the uncertainty on $T^*$ to about 20 K. Within uncertainties, the values for the critical temperature of the crossover obtained from ARPES and from the Knight shift are thus consistent.  

In the framework of DMFT calculations, electronic correlations are predicted to increase as we approach the half-filling of the $d$ electronic shell or a subset of the $d$ bands isolated in energy \cite{HauleNJP11,WernerNatPhys8,deMedici_PRL107,vanRoekeghemCR17}. This concept was successfully tested experimentally in the 122 family of 3$d$ transition metal pnictides, where the band renormalization measured by ARPES increases monotonically from BaCu$_2$As$_2$ (3$d^{10}$) to BaFe$_2$As$_2$ (3$d^6$) \cite{LK_ZengPRB94}. The complete substitution of Ba by K in BaFe$_2$As$_2$ corresponds to an additional nominal decrease of 0.5 electron per Fe, and thus stronger correlations are expected for the latter material \cite{WernerNatPhys8}. Although local electronic correlations seem \textit{a priori} a good candidate for explaining the
crossover in the electrical resistivity and the Knight shift, it is worth recalling that the $\beta$ band with $d_{xy}$ character, which ARPES measurements on the 122 ferropnictide materials demonstrated to be more correlated than the others by a factor of 2 \cite{Nan_XuPRX3}, is already very incoherent at 20 K in KFe$_2$As$_2$, way below the crossover temperature. Moreover, the fact
that the loss of coherence can be fitted by an exponential function suggests that the loss of coherence may come from a thermal process.  

In order to find alternative explanations for the coherent to incoherent crossover, in Figs. \ref{Tcuts}(a)-\ref{Tcuts}(j) we show the temperature evolution of the KFe$_2$As$_2$ spectra from 20 to 200 K.  To minimize extrinsic effects related to the Fermi-Dirac cutoff, we divide all the spectra by the F$_{\textrm{D}}$ function convoluted by the instrumental resolution function. As temperature increases and the F$_{\textrm{D}}$ function broadens, we can access partly the band structure above $E_F$. In particular, the data reveal a very complex electronic structure at the M point. Together, the $\varepsilon$ and $\delta$ bands form a ``M"-shape feature toping only a few meV above $E_F$. At this top, located around $0.2\pi/a$ away from the M point, the hole-like $\varepsilon$ and electron-like $\delta$ band dispersions hybridize and open a very small gap. Although we cannot resolve the portion of the $\delta$ band above the ``M"-shape feature, the continuation of the $\varepsilon$ band is clearly visible and tops slightly above $E_F$. 

Interestingly, our analysis suggests that the ``M"-shape feature shifts downwards with increasing temperature. From 20 K to 120 K, temperature above which it becomes difficult to trace the band dispersions precisely, this shift is about 7.5 meV, as illustrated in Fig. \ref{Tcuts}(k). As indicated in Fig. \ref{Tcuts}(k), a similar downward shift of 9 meV is observed for the $\alpha$ band. Such shifts of the band structure are not unique to KFe$_2$As$_2$ and stronger temperature effects have already been reported in Ba(Fe$_{1-x}$Co$_x$)$_2$As$_2$ \cite{Brouet_PRL110}, Ba(Fe$_{1-x}$Ru$_x$)$_2$As$_2$ \cite{Dhaka_PRL110} and Fe$_{1.06}$Te \cite{PH_LinPRL111}. For a large part, these shifts are induced by the occupied-unoccupied asymmetry in the band structure. 

The consequence of this shift in KFe$_2$As$_2$ may be important since the data suggest that the $\varepsilon$ pocket possibly sinks below $E_F$ at some temperature above 120 K. Although the broadening of the Fermi-Dirac function at such high temperature prevents us from calling this phenomenon a Lifshitz transition \cite{Lifshitz_JETP11}, one should not neglect the impact on the electronic transport properties that may have the large density-of-states accompanying a saddle point, as well as the top or the bottom of a band located slightly below or above $E_F$ \cite{FinkEPL113,Drechsler_PSSB254}. Notably, recently we identified a van Hove singularity near $E_F$ in KFe$_2$As$_2$ \cite{DL_Fang_vHs}, which we showed in Section \ref{Cs_section} to contribute to the large Sommerfeld coefficient of this material \cite{YP_Wu_PRL116} and for the sudden reversal in the pressure dependence of the superconducting transition temperature reported in AFe$_2$As$_2$ (A = K, Rb, Cs) \cite{Tafti_nphys9,Tafti_PRB89,Tafti_PRB91,Wiecki_PRB97}. Could the van Hove singularity be also the origin of the crossover at $T^*$? 

The behavior of $T^*$ under pressure or following the chemical substitution of K by Rb or Cs is qualitatively consistent with the assumption that the van Hove singularity determines $T^*$, thus challenging previous interpretations of $T^*$ \cite{Hardy_PRL111,YP_Wu_PRL116}. Within the van Hove singularity scenario, the closer the van Hove singularity to $E_F$, the lower $T^*$ should be because the large density-of-states accompanying the van Hove singularity is easier to activate thermally. This effect seems to be captured by very recent calculations of the spin susceptibility for a one-band Hubbard model near half-filling \cite{Nourafkan_Knight_shift}. The decrease of $T^*$ under substitution of K by Cs \cite{Wiecki_PRB97}, \textit{i.~e.} with increasing chemical pressure, is consistent with a shift of the van Hove singularity towards lower binding energies (see Section \ref{Cs_section}). In KFe$_2$As$_2$, $T^*$ increases with applied pressure \cite{Wiecki_PRB97}, suggesting a shift of the van Hove singularity towards higher binding energies also consistent with the effect of chemical pressure (see Section \ref{Cs_section}). We point out that the same reasoning is also valid for the top and bottom of bands found within about 30 meV above $E_F$. It is also important to note that although the small temperature shift of the van Hove singularity away from $E_F$ that we reported above would contribute to elevate $T^*$, the shift is smaller than the thermal broadening.

\begin{figure}[!t]
\begin{center}
\includegraphics[width=\columnwidth]{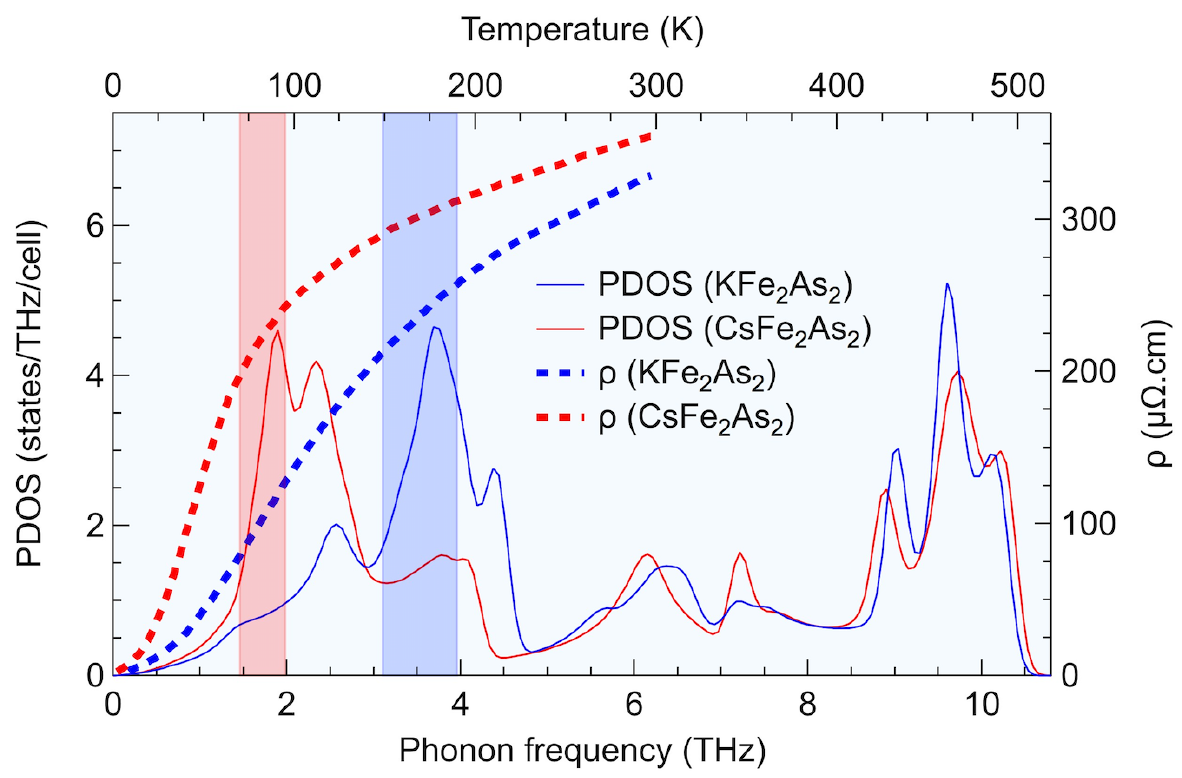}
\end{center}
\caption{\label{PDOS}(Color online) PDOS of KFe$_2$As$_2$ and CsFe$_2$As$_2$ (solid lines), plotted using the left axis as a function of frequency (bottom axis). The dashed lines refer to the corresponding resistivity curves extracted from Ref. \cite{YP_Wu_PRL116}, and plotted using the right axis as a function of temperature (top axis). The shadow areas are also extracted from Ref. \cite{YP_Wu_PRL116} and correspond to the ranges of the coherence temperature crossovers of KFe$_2$As$_2$ (blue) and CsFe$_2$As$_2$ (red).} 
\end{figure}

Another candidate for explaining the anomaly at $T^*$ are phonons. This possibility has been mentioned in Ref. \cite{Wiecki_PRB97}, which suggests that the decrease in resistivity could in principle be due to a small Debye temperature. In order to investigate this scenario, we calculated the PDOS of KFe$_2$As$_2$ and CsFe$_2$As$_2$ within the LDA at room temperature, including anharmonic effects as described in Ref. \cite{vanRoekeghem_PRB94}, using a $4\times 4\times 1$ supercell of the tetragonal paramagnetic phase. Using the proper scalings for the equivalence between frequency and temperature (1~THz~=~47.99243~K), in Fig. \ref{PDOS} we overlap the PDOS of KFe$_2$As$_2$ and CsFe$_2$As$_2$ with the corresponding resistivity curves extracted from Ref. \cite{YP_Wu_PRL116}. Interestingly, the energy of the first main peak of the PDOS coincides with the coherence crossover temperature obtained from the resistivity data. This peak is mainly linked to states involving the cation (K or Cs), such that we expect that the PDOS of RbFe$_2$As$_2$ would also correspond to the $T^*$ value reported for that compound \cite{YP_Wu_PRL116}, simply based on the different atomic masses of K, Rb and Cs. As shown in Ref. \cite{Allan_PRB17}, it is then expected that above $T^*$ the resistivity will start to behave as a linear function of temperature due to electron-phonon scattering. We also speculate that this phenomenon might explain the linear resistivity trends at high temperatures observed in many Fe pnictides and discussed at length in the literature. For instance, in the CaFe$_2$As$_2$ family, the change of resistivity from linear to quadratic behavior after the collapse transition might well result from the shift of the phonon spectrum towards higher frequencies \cite{Mittal_PRB81}, while the associated reduction of electronic correlations is relatively small and mainly due to in-plane structural changes rather than out-of-plane parameters \cite{vanRoekeghem_PRB93}.

\section{Discussion}

In this paper, we present evidence supporting the idea that the large enhancement of the effective mass in KFe$_2$As$_2$ and CsFe$_2$As$_2$ as compared to optimally-doped Ba$_{1-x}$K$_{x}$Fe$_2$As$_2$ is not caused by enhanced electronic Coulomb correlations due to the presence of large Hund's coupling alone, but that the existence of a van Hove singularity in the vicinity of $E_F$ contributes significantly to the large Sommerfeld coefficient. We caution that this does not mean that KFe$_2$As$_2$ is uncorrelated. On the contrary, the faint $\beta$ band is a good indication that the system is correlated. Similarly, while our results indicate that the van Hove singularity and its variation of position can trigger a change in the superconducting order parameter, electronic correlations are still a likely candidate for the driving force of Cooper pairing.  Along with spin-orbit coupling \cite{Drechsler_PSSB254}, electronic correlations beyond the simple density functional picture should also determine the precise location of the van Hove singularity.

We discuss the temperature evolution of the electronic properties in terms
of {\it two} crossover temperatures, a low-energy scale of the order of a
few tens of Kelvin, where Fermi liquid coherence is lost, and a high-energy scale identified previously from NMR measurements and related to
a change between two different rather linear regimes in the resistivity.
The latter energy scale coincides with the energy scales of pronounced
first peaks in the phonon densities of modes for KFe$_2$As$_2$ and
CsFe$_2$As$_2$.
\vspace{0.3cm}

\section{Summary}

In summary, we compared the electronic structure of KFe$_2$As$_2$ and CsFe$_2$As$_2$ using ARPES and first-principles calculations. We show that although the Fermi surfaces of these two materials are very similar, the electronic structure of KFe$_2$As$_2$ is more three-dimensional. Notably, a van Hove singularity previously reported in the vicinity of $E_F$ in KFe$_2$As$_2$ moves even closer to $E_F$ upon negative chemical pressure, resulting in a larger Sommerfeld coefficient due to the increase in the density-of-states. This van Hove singularity is the most likely candidate to explain the high-energy crossover in the resisitivity 
the temperature of which correlates with the sudden reversal in the pressure dependence of $T_c$. The momentum location of the van Hove singularity on the nodal line of the $s_{\pm}$ gap function, very near the $\varepsilon$ FS, offers a possible explanation for this behavior, as it moves closer or away to $E_F$. We also showed that the spectral weight decreases exponentially with a characteristic temperature $T^*$ consistent with values reported previously using different experimental techniques. We showed that the low-energy electronic structure of KFe$_2$As$_2$ with a van Hove singularity, combined with a peaked PDOS can explain the high-energy crossover without explicit need for strong electronic correlations. 

\section*{Acknowledgments}

We thank Reza Nourafkan, Andr\'{e}-Marie Tremblay and Louis Taillefer for useful discussions. This work was supported by grants from NSFC (Grants Nos. 11274362 and 11674371) and MOST (Grants Nos. 2015CB921301, 2016YFA0300300 and 2016YFA0401000) from China, a Consolidator Grant of the European Research Council (Project No. 617196) and supercomputing time at IDRIS/GENCI Orsay (Project No. t2018091393). This research was undertaken thanks in part to funding from the Canada First Research Excellence Fund. We acknowledge Diamond Light Source for time on beamline I05 under proposals SI9469 and SI11452, which contributed to the results presented here.

\bibliography{biblio_long_titles}

\begin{thebibliography}{61}%
\makeatletter
\providecommand \@ifxundefined [1]{%
 \@ifx{#1\undefined}
}%
\providecommand \@ifnum [1]{%
 \ifnum #1\expandafter \@firstoftwo
 \else \expandafter \@secondoftwo
 \fi
}%
\providecommand \@ifx [1]{%
 \ifx #1\expandafter \@firstoftwo
 \else \expandafter \@secondoftwo
 \fi
}%
\providecommand \natexlab [1]{#1}%
\providecommand \enquote  [1]{``#1''}%
\providecommand \bibnamefont  [1]{#1}%
\providecommand \bibfnamefont [1]{#1}%
\providecommand \citenamefont [1]{#1}%
\providecommand \href@noop [0]{\@secondoftwo}%
\providecommand \href [0]{\begingroup \@sanitize@url \@href}%
\providecommand \@href[1]{\@@startlink{#1}\@@href}%
\providecommand \@@href[1]{\endgroup#1\@@endlink}%
\providecommand \@sanitize@url [0]{\catcode `\\12\catcode `\$12\catcode
  `\&12\catcode `\#12\catcode `\^12\catcode `\_12\catcode `\%12\relax}%
\providecommand \@@startlink[1]{}%
\providecommand \@@endlink[0]{}%
\providecommand \url  [0]{\begingroup\@sanitize@url \@url }%
\providecommand \@url [1]{\endgroup\@href {#1}{\urlprefix }}%
\providecommand \urlprefix  [0]{URL }%
\providecommand \Eprint [0]{\href }%
\providecommand \doibase [0]{http://dx.doi.org/}%
\providecommand \selectlanguage [0]{\@gobble}%
\providecommand \bibinfo  [0]{\@secondoftwo}%
\providecommand \bibfield  [0]{\@secondoftwo}%
\providecommand \translation [1]{[#1]}%
\providecommand \BibitemOpen [0]{}%
\providecommand \bibitemStop [0]{}%
\providecommand \bibitemNoStop [0]{.\EOS\space}%
\providecommand \EOS [0]{\spacefactor3000\relax}%
\providecommand \BibitemShut  [1]{\csname bibitem#1\endcsname}%
\let\auto@bib@innerbib\@empty
\bibitem [{\citenamefont {{P. Werner, E. Gull, M. Troyer and A. J.
  Millis}}(2008)}]{WernerPRL101}%
  \BibitemOpen
  \bibfield  {author} {\bibinfo {author} {\bibnamefont {{P. Werner, E. Gull, M.
  Troyer and A. J. Millis}}},\ }\bibfield  {title} {\enquote {\bibinfo {title}
  {{Spin Freezing Transition and Non-Fermi-Liquid Self-Energy in a
  Three-Orbital Model}},}\ }\href@noop {} {\bibfield  {journal} {\bibinfo
  {journal} {Phys. Rev. Lett.}\ }\textbf {\bibinfo {volume} {101}},\ \bibinfo
  {pages} {166405} (\bibinfo {year} {2008})}\BibitemShut {NoStop}%
\bibitem [{\citenamefont {{K. Haule and G. Kotliar}}(2009)}]{HauleNJP11}%
  \BibitemOpen
  \bibfield  {author} {\bibinfo {author} {\bibnamefont {{K. Haule and G.
  Kotliar}}},\ }\bibfield  {title} {\enquote {\bibinfo {title}
  {{Coherence-incoherence crossover in the normal state of iron oxypnictides
  and importance of Hund's rule coupling}},}\ }\href@noop {} {\bibfield
  {journal} {\bibinfo  {journal} {New J. Phys.}\ }\textbf {\bibinfo {volume}
  {11}},\ \bibinfo {pages} {025021} (\bibinfo {year} {2009})}\BibitemShut
  {NoStop}%
\bibitem [{\citenamefont {{P. Werner, M. Casula, T. Miyake, F. Aryasetiawan, A.
  J. Millis and S. Biermann}}(2012)}]{WernerNatPhys8}%
  \BibitemOpen
  \bibfield  {author} {\bibinfo {author} {\bibnamefont {{P. Werner, M. Casula,
  T. Miyake, F. Aryasetiawan, A. J. Millis and S. Biermann}}},\ }\bibfield
  {title} {\enquote {\bibinfo {title} {{Satellites and large doping and
  temperature dependence of electronic properties in hole-doped
  BaFe$_2$As$_2$}},}\ }\href@noop {} {\bibfield  {journal} {\bibinfo  {journal}
  {Nature Phys.}\ }\textbf {\bibinfo {volume} {8}},\ \bibinfo {pages} {331}
  (\bibinfo {year} {2012})}\BibitemShut {NoStop}%
\bibitem [{\citenamefont {{L. de' Medici, J. Mravlje and A.
  Georges}}(2011)}]{deMedici_PRL107}%
  \BibitemOpen
  \bibfield  {author} {\bibinfo {author} {\bibnamefont {{L. de' Medici, J.
  Mravlje and A. Georges}}},\ }\bibfield  {title} {\enquote {\bibinfo {title}
  {{Janus-Faced Influence of Hund's Rule Coupling in Strongly Correlated
  Materials}},}\ }\href@noop {} {\bibfield  {journal} {\bibinfo  {journal}
  {Phys. Rev. Lett.}\ }\textbf {\bibinfo {volume} {107}},\ \bibinfo {pages}
  {256401} (\bibinfo {year} {2011})}\BibitemShut {NoStop}%
\bibitem [{\citenamefont {{Ambroise van Roekeghem, Pierre Richard, Hong Ding
  and Silke Biermann }}(2015)}]{vanRoekeghemCR17}%
  \BibitemOpen
  \bibfield  {author} {\bibinfo {author} {\bibnamefont {{Ambroise van
  Roekeghem, Pierre Richard, Hong Ding and Silke Biermann }}},\ }\bibfield
  {title} {\enquote {\bibinfo {title} {{Spectral properties of transition metal
  pnictides and chalcogenides: Angle-resolved photoemission spectroscopy and
  dynamical mean-field theory}},}\ }\href@noop {} {\bibfield  {journal}
  {\bibinfo  {journal} {C. R. Physique}\ }\textbf {\bibinfo {volume} {17}},\
  \bibinfo {pages} {140} (\bibinfo {year} {2015})}\BibitemShut {NoStop}%
\bibitem [{\citenamefont {{F. Hardy, A.E. B\"{o}hmer, D. Aoki, P. Burger, T.
  Wolf, P. Schweiss, R. Heid, P. Adelmann, Y. X. Yao, G. Kotliar, J. Schmalian
  and C. Meingast}}(2013)}]{Hardy_PRL111}%
  \BibitemOpen
  \bibfield  {author} {\bibinfo {author} {\bibnamefont {{F. Hardy, A.E.
  B\"{o}hmer, D. Aoki, P. Burger, T. Wolf, P. Schweiss, R. Heid, P. Adelmann,
  Y. X. Yao, G. Kotliar, J. Schmalian and C. Meingast}}},\ }\bibfield  {title}
  {\enquote {\bibinfo {title} {{Evidence of Strong Correlations and
  Coherence-Incoherence Crossover in the Iron Pnictide Superconductor
  KFe$_2$As$_2$}},}\ }\href@noop {} {\bibfield  {journal} {\bibinfo  {journal}
  {Phys. Rev. Lett.}\ }\textbf {\bibinfo {volume} {111}},\ \bibinfo {pages}
  {027002} (\bibinfo {year} {2013})}\BibitemShut {NoStop}%
\bibitem [{\citenamefont {{Y. P. Wu, D. Zhao, A. F. Wang, N. Z. Wang, Z. J.
  Xiang, X. G. Luo, T. Wu and X. H. Chen}}(2016)}]{YP_Wu_PRL116}%
  \BibitemOpen
  \bibfield  {author} {\bibinfo {author} {\bibnamefont {{Y. P. Wu, D. Zhao, A.
  F. Wang, N. Z. Wang, Z. J. Xiang, X. G. Luo, T. Wu and X. H. Chen}}},\
  }\bibfield  {title} {\enquote {\bibinfo {title} {{Emergent Kondo Lattice
  Behavior in Iron-Based Superconductors $A$Fe$_2$As$_2$ (A = K, Rb, Cs)}},}\
  }\href@noop {} {\bibfield  {journal} {\bibinfo  {journal} {Phys. Rev. Lett.}\
  }\textbf {\bibinfo {volume} {116}},\ \bibinfo {pages} {147001} (\bibinfo
  {year} {2016})}\BibitemShut {NoStop}%
\bibitem [{\citenamefont {{P. Wiecki, V. Taufour, D. Y. Chung, M. G.
  Kanatzidis, S. L. Bud'ko, P. C. Canfield and Y.
  Furukawa}}(2018)}]{Wiecki_PRB97}%
  \BibitemOpen
  \bibfield  {author} {\bibinfo {author} {\bibnamefont {{P. Wiecki, V. Taufour,
  D. Y. Chung, M. G. Kanatzidis, S. L. Bud'ko, P. C. Canfield and Y.
  Furukawa}}},\ }\bibfield  {title} {\enquote {\bibinfo {title} {{Pressure
  dependence of coherence-incoherence crossover behavior in KFe$_2$As$_2$
  observed by resistivity and $^{75}$As-NMR/NQR}},}\ }\href@noop {} {\bibfield
  {journal} {\bibinfo  {journal} {Phys. Rev. B}\ }\textbf {\bibinfo {volume}
  {97}},\ \bibinfo {pages} {064509} (\bibinfo {year} {2018})}\BibitemShut
  {NoStop}%
\bibitem [{\citenamefont {{K. Okazaki, Y. Ota, Y. Kotani, W. Malaeb, Y. Ishida,
  T. Shimojima, T. Kiss, S. Watanabe, C.-T. Chen, K. Kihou, C. H. Lee, A. Iyo,
  H. Eisaki, T. Saito, H. Fukazawa, Y. Kohori, K. Hashimoto, T. Shibauchi
  \textit{et al.}}}(2012)}]{Okazaki_Science337}%
  \BibitemOpen
  \bibfield  {author} {\bibinfo {author} {\bibnamefont {{K. Okazaki, Y. Ota, Y.
  Kotani, W. Malaeb, Y. Ishida, T. Shimojima, T. Kiss, S. Watanabe, C.-T. Chen,
  K. Kihou, C. H. Lee, A. Iyo, H. Eisaki, T. Saito, H. Fukazawa, Y. Kohori, K.
  Hashimoto, T. Shibauchi \textit{et al.}}}},\ }\bibfield  {title} {\enquote
  {\bibinfo {title} {{Octet-Line Node Structure of Superconducting Order
  Parameter in KFe$_2$As$_2$}},}\ }\href@noop {} {\bibfield  {journal}
  {\bibinfo  {journal} {Science}\ }\textbf {\bibinfo {volume} {337}},\ \bibinfo
  {pages} {1314} (\bibinfo {year} {2012})}\BibitemShut {NoStop}%
\bibitem [{\citenamefont {{J. K. Dong, S.Y. Zhou, T.Y. Guan, H. Zhang, Y. F.
  Dai, X. Qiu, X. F. Wang, Y. He, X. H. Chen and S.Y.
  Li}}(2010)}]{JK_DongPRL2010}%
  \BibitemOpen
  \bibfield  {author} {\bibinfo {author} {\bibnamefont {{J. K. Dong, S.Y. Zhou,
  T.Y. Guan, H. Zhang, Y. F. Dai, X. Qiu, X. F. Wang, Y. He, X. H. Chen and
  S.Y. Li}}},\ }\bibfield  {title} {\enquote {\bibinfo {title} {{Quantum
  Criticality and Nodal Superconductivity in the FeAs-Based Superconductor
  KFe$_2$As$_2$}},}\ }\href@noop {} {\bibfield  {journal} {\bibinfo  {journal}
  {Phys. Rev. Lett.}\ }\textbf {\bibinfo {volume} {104}},\ \bibinfo {pages}
  {087005} (\bibinfo {year} {2010})}\BibitemShut {NoStop}%
\bibitem [{\citenamefont {{J.-Ph. Reid, M. A. Tanatar, A. Juneau-Fecteau, R. T.
  Gordon, S. Ren\'{e} de Cotret, N. Doiron-Leyraud, T. Saito, H. Fukazawa, Y.
  Kohori, K. Kihou, C. H. Lee, A. Iyo, H. Eisaki, R. Prozorov and L.
  Taillefer}}(2012)}]{ReidPRL109}%
  \BibitemOpen
  \bibfield  {author} {\bibinfo {author} {\bibnamefont {{J.-Ph. Reid, M. A.
  Tanatar, A. Juneau-Fecteau, R. T. Gordon, S. Ren\'{e} de Cotret, N.
  Doiron-Leyraud, T. Saito, H. Fukazawa, Y. Kohori, K. Kihou, C. H. Lee, A.
  Iyo, H. Eisaki, R. Prozorov and L. Taillefer}}},\ }\bibfield  {title}
  {\enquote {\bibinfo {title} {{Universal Heat Conduction in the Iron Arsenide
  Superconductor KFe$_2$As$_2$: Evidence of a $d$-Wave State}},}\ }\href@noop
  {} {\bibfield  {journal} {\bibinfo  {journal} {Phys. Rev. Lett.}\ }\textbf
  {\bibinfo {volume} {109}},\ \bibinfo {pages} {087001} (\bibinfo {year}
  {2012})}\BibitemShut {NoStop}%
\bibitem [{\citenamefont {{X. C. Hong, X. L. Li, B. Y. Pan, L. P. He, A. F.
  Wang, X. G. Luo, X. H. Chen and S. Y. Li}}(2014)}]{XC_HongPRB87}%
  \BibitemOpen
  \bibfield  {author} {\bibinfo {author} {\bibnamefont {{X. C. Hong, X. L. Li,
  B. Y. Pan, L. P. He, A. F. Wang, X. G. Luo, X. H. Chen and S. Y. Li}}},\
  }\bibfield  {title} {\enquote {\bibinfo {title} {{Nodal gap in iron-based
  superconductor CsFe$_2$As$_2$ probed by quasiparticle heat transport}},}\
  }\href@noop {} {\bibfield  {journal} {\bibinfo  {journal} {Phys. Rev. B}\
  }\textbf {\bibinfo {volume} {87}},\ \bibinfo {pages} {144502} (\bibinfo
  {year} {2014})}\BibitemShut {NoStop}%
\bibitem [{\citenamefont {{A. F. Wang, S. Y. Zhou, X. G. Luo, X. C. Hong, Y. J.
  Yan, J. J. Ying, P. Cheng, G. J. Ye, Z. J. Xiang, S. Y. Li and X. H.
  Chen}}(2014)}]{AF_WangPRB89}%
  \BibitemOpen
  \bibfield  {author} {\bibinfo {author} {\bibnamefont {{A. F. Wang, S. Y.
  Zhou, X. G. Luo, X. C. Hong, Y. J. Yan, J. J. Ying, P. Cheng, G. J. Ye, Z. J.
  Xiang, S. Y. Li and X. H. Chen}}},\ }\bibfield  {title} {\enquote {\bibinfo
  {title} {{Anomalous impurity effects in the iron-based superconductor
  KFe$_2$As$_2$}},}\ }\href@noop {} {\bibfield  {journal} {\bibinfo  {journal}
  {Phys. Rev. B}\ }\textbf {\bibinfo {volume} {89}},\ \bibinfo {pages} {064510}
  (\bibinfo {year} {2014})}\BibitemShut {NoStop}%
\bibitem [{\citenamefont {{Z. Zhang, A. F. Wang, X. C. Hong, J. Zhang, B. Y.
  Pan, J. Pan, Y. Xu, X. G. Luo, X. H. Chen and S. Y.
  Li}}(2015)}]{Z_Zhang_PRB91}%
  \BibitemOpen
  \bibfield  {author} {\bibinfo {author} {\bibnamefont {{Z. Zhang, A. F. Wang,
  X. C. Hong, J. Zhang, B. Y. Pan, J. Pan, Y. Xu, X. G. Luo, X. H. Chen and S.
  Y. Li}}},\ }\bibfield  {title} {\enquote {\bibinfo {title} {{Heat transport
  in RbFe$_2$As$_2$ single crystals: Evidence for nodal superconducting
  gap}},}\ }\href@noop {} {\bibfield  {journal} {\bibinfo  {journal} {Phys.
  Rev. B}\ }\textbf {\bibinfo {volume} {91}},\ \bibinfo {pages} {024502}
  (\bibinfo {year} {2015})}\BibitemShut {NoStop}%
\bibitem [{\citenamefont {{M. Abdel-Hafiez, S. Aswartham, S. Wurmehl, V.
  Grinenko, C. Hess, S.-L. Drechsler, S. Johnston, A. U. B. Wolter, B.
  B\"{u}chner, H. Rosner and L. Boeri}}(2012)}]{Abdel-HafiezPRB85}%
  \BibitemOpen
  \bibfield  {author} {\bibinfo {author} {\bibnamefont {{M. Abdel-Hafiez, S.
  Aswartham, S. Wurmehl, V. Grinenko, C. Hess, S.-L. Drechsler, S. Johnston, A.
  U. B. Wolter, B. B\"{u}chner, H. Rosner and L. Boeri}}},\ }\bibfield  {title}
  {\enquote {\bibinfo {title} {{Specific heat and upper critical fields in
  KFe$_2$As$_2$ single crystals}},}\ }\href@noop {} {\bibfield  {journal}
  {\bibinfo  {journal} {Phys. Rev. B}\ }\textbf {\bibinfo {volume} {85}},\
  \bibinfo {pages} {134533} (\bibinfo {year} {2012})}\BibitemShut {NoStop}%
\bibitem [{\citenamefont {{A. F. Wang, B. Y. Pan, X. G. Luo, F. Chen, Y. J.
  Yan, J. J. Ying, G. J. Ye, P. Cheng, X. C. Hong, S. Y. Li and X. H.
  Chen}}(2013)}]{AF_WangPRB87}%
  \BibitemOpen
  \bibfield  {author} {\bibinfo {author} {\bibnamefont {{A. F. Wang, B. Y. Pan,
  X. G. Luo, F. Chen, Y. J. Yan, J. J. Ying, G. J. Ye, P. Cheng, X. C. Hong, S.
  Y. Li and X. H. Chen}}},\ }\bibfield  {title} {\enquote {\bibinfo {title}
  {{Calorimetric study of single-crystal CsFe$_2$As$_2$}},}\ }\href@noop {}
  {\bibfield  {journal} {\bibinfo  {journal} {Phys. Rev. B}\ }\textbf {\bibinfo
  {volume} {87}},\ \bibinfo {pages} {214509} (\bibinfo {year}
  {2013})}\BibitemShut {NoStop}%
\bibitem [{\citenamefont {{S. Kittaka, Y. Aoki, N. Kase, T. Sakakibara, T.
  Saito, H. Fukazawa, Y. Kohori, K. Kihou, C.-H. Lee, A. Iyo, H. Eisaki, K.
  Deguchi, N. K. Sato, Y. Tsutsumi and K. Machida}}(2014)}]{KittakaJPSJ2014}%
  \BibitemOpen
  \bibfield  {author} {\bibinfo {author} {\bibnamefont {{S. Kittaka, Y. Aoki,
  N. Kase, T. Sakakibara, T. Saito, H. Fukazawa, Y. Kohori, K. Kihou, C.-H.
  Lee, A. Iyo, H. Eisaki, K. Deguchi, N. K. Sato, Y. Tsutsumi and K.
  Machida}}},\ }\bibfield  {title} {\enquote {\bibinfo {title} {{Thermodynamic
  Study of Nodal Structure and Multiband Superconductivity of
  KFe$_2$As$_2$}},}\ }\href@noop {} {\bibfield  {journal} {\bibinfo  {journal}
  {J. Phys. Soc. Jpn.}\ }\textbf {\bibinfo {volume} {83}},\ \bibinfo {pages}
  {013704} (\bibinfo {year} {2014})}\BibitemShut {NoStop}%
\bibitem [{\citenamefont {{K. Hashimoto, A. Serafin, S. Tonegawa, R. Katsumata,
  R. Okazaki, T. Saito, H. Fukazawa, Y. Kohori, K. Kihou, C. H. Lee, A. Iyo, H.
  Eisaki, H. Ikeda, Y. Matsuda, A. Carrington and T.
  Shibauchi}}(2010)}]{Hashimoto_PRB82}%
  \BibitemOpen
  \bibfield  {author} {\bibinfo {author} {\bibnamefont {{K. Hashimoto, A.
  Serafin, S. Tonegawa, R. Katsumata, R. Okazaki, T. Saito, H. Fukazawa, Y.
  Kohori, K. Kihou, C. H. Lee, A. Iyo, H. Eisaki, H. Ikeda, Y. Matsuda, A.
  Carrington and T. Shibauchi}}},\ }\bibfield  {title} {\enquote {\bibinfo
  {title} {{Evidence for superconducting gap nodes in the zone-centered hole
  bands of KFe$_2$As$_2$ from magnetic penetration-depth measurements}},}\
  }\href@noop {} {\bibfield  {journal} {\bibinfo  {journal} {Phys. Rev. B}\
  }\textbf {\bibinfo {volume} {82}},\ \bibinfo {pages} {014526} (\bibinfo
  {year} {2010})}\BibitemShut {NoStop}%
\bibitem [{\citenamefont {{Y. Mizukami, Y. Kawamoto, Y. Shimoyama, S. Kurata,
  H. Ikeda, T. Wolf, D. A. Zocco, K. Grube, H. v. L\"{o}hneysen, Y. Matsuda and
  T. Shibauchi}}(2016)}]{Mizukami_PRB89}%
  \BibitemOpen
  \bibfield  {author} {\bibinfo {author} {\bibnamefont {{Y. Mizukami, Y.
  Kawamoto, Y. Shimoyama, S. Kurata, H. Ikeda, T. Wolf, D. A. Zocco, K. Grube,
  H. v. L\"{o}hneysen, Y. Matsuda and T. Shibauchi}}},\ }\bibfield  {title}
  {\enquote {\bibinfo {title} {{Evolution of quasiparticle excitations with
  enhanced electron correlations in superconducting AFe$_2$As$_2$ (A = K, Rb,
  and Cs)}},}\ }\href@noop {} {\bibfield  {journal} {\bibinfo  {journal} {Phys.
  Rev. B}\ }\textbf {\bibinfo {volume} {94}},\ \bibinfo {pages} {024508}
  (\bibinfo {year} {2016})}\BibitemShut {NoStop}%
\bibitem [{\citenamefont {{F. F. Tafti, A. Juneau-Fecteau, M-\`{E}. Delage, S.
  Ren\'{e} de Cotret, J-Ph. Reid, A. F. Wang, X-G. Luo, X. H. Chen, N.
  Doiron-Leyraud and L. Taillefer}}(2013)}]{Tafti_nphys9}%
  \BibitemOpen
  \bibfield  {author} {\bibinfo {author} {\bibnamefont {{F. F. Tafti, A.
  Juneau-Fecteau, M-\`{E}. Delage, S. Ren\'{e} de Cotret, J-Ph. Reid, A. F.
  Wang, X-G. Luo, X. H. Chen, N. Doiron-Leyraud and L. Taillefer}}},\
  }\bibfield  {title} {\enquote {\bibinfo {title} {{Sudden reversal in the
  pressure dependence of $T_c$ in the iron-based superconductor
  KFe$_2$As$_2$}},}\ }\href@noop {} {\bibfield  {journal} {\bibinfo  {journal}
  {Nature Phys.}\ }\textbf {\bibinfo {volume} {9}},\ \bibinfo {pages} {349}
  (\bibinfo {year} {2013})}\BibitemShut {NoStop}%
\bibitem [{\citenamefont {{F. F. Tafti, J. P. Clancy, M. Lapointe-Major, C.
  Collignon, S. Faucher, J. A. Sears, A. Juneau-Fecteau, N. Doiron-Leyraud, A.
  F. Wang, X.-G. Luo, X. H. Chen, S. Desgreniers, Young-June Kim, and Louis
  Taillefer}}(2014)}]{Tafti_PRB89}%
  \BibitemOpen
  \bibfield  {author} {\bibinfo {author} {\bibnamefont {{F. F. Tafti, J. P.
  Clancy, M. Lapointe-Major, C. Collignon, S. Faucher, J. A. Sears, A.
  Juneau-Fecteau, N. Doiron-Leyraud, A. F. Wang, X.-G. Luo, X. H. Chen, S.
  Desgreniers, Young-June Kim, and Louis Taillefer}}},\ }\bibfield  {title}
  {\enquote {\bibinfo {title} {{Sudden reversal in the pressure dependence of
  $T_c$ in the iron-based superconductor CsFe$_2$As$_2$: A possible link
  between inelastic scattering and pairing symmetry}},}\ }\href@noop {}
  {\bibfield  {journal} {\bibinfo  {journal} {Phys. Rev. B}\ }\textbf {\bibinfo
  {volume} {89}},\ \bibinfo {pages} {134502} (\bibinfo {year}
  {2014})}\BibitemShut {NoStop}%
\bibitem [{\citenamefont {{F. F. Tafti, A. Ouellet, A. Juneau-Fecteau, S.
  Faucher, M. Lapointe-Major, N. Doiron-Leyraud, A. F. Wang, X.-G. Luo, X. H.
  Chen and Louis Taillefer}}(2015)}]{Tafti_PRB91}%
  \BibitemOpen
  \bibfield  {author} {\bibinfo {author} {\bibnamefont {{F. F. Tafti, A.
  Ouellet, A. Juneau-Fecteau, S. Faucher, M. Lapointe-Major, N. Doiron-Leyraud,
  A. F. Wang, X.-G. Luo, X. H. Chen and Louis Taillefer}}},\ }\bibfield
  {title} {\enquote {\bibinfo {title} {{Universal V-shaped temperature-pressure
  phase diagram in the iron-based superconductors KFe$_2$As$_2$,
  RbFe$_2$As$_2$, and CsFe$_2$As$_2$}},}\ }\href@noop {} {\bibfield  {journal}
  {\bibinfo  {journal} {Phys. Rev. B}\ }\textbf {\bibinfo {volume} {91}},\
  \bibinfo {pages} {054511} (\bibinfo {year} {2015})}\BibitemShut {NoStop}%
\bibitem [{\citenamefont {{P. S. Wang, P. Zhou, J. Dai, J. Zhang, X. X. Ding,
  H. Lin, H. H. Wen, B. Normand, R. Yu and Weiqiang Yu}}(2016)}]{PS_WangPRB93}%
  \BibitemOpen
  \bibfield  {author} {\bibinfo {author} {\bibnamefont {{P. S. Wang, P. Zhou,
  J. Dai, J. Zhang, X. X. Ding, H. Lin, H. H. Wen, B. Normand, R. Yu and
  Weiqiang Yu}}},\ }\bibfield  {title} {\enquote {\bibinfo {title} {{Nearly
  critical spin and charge fluctuations in KFe$_2$As$_2$ observed by
  high-pressure NMR}},}\ }\href@noop {} {\bibfield  {journal} {\bibinfo
  {journal} {Phys. Rev. B}\ }\textbf {\bibinfo {volume} {93}},\ \bibinfo
  {pages} {085129} (\bibinfo {year} {2016})}\BibitemShut {NoStop}%
\bibitem [{\citenamefont {{Diego A. Zocco, Kai Grube, Felix Eilers, Thomas Wolf
  and Hilbert v. L\"{o}hneysen}}(2014)}]{JPSConf2014_Zocco}%
  \BibitemOpen
  \bibfield  {author} {\bibinfo {author} {\bibnamefont {{Diego A. Zocco, Kai
  Grube, Felix Eilers, Thomas Wolf and Hilbert v. L\"{o}hneysen}}},\ }\bibfield
   {title} {\enquote {\bibinfo {title} {{Fermi Surface of KFe$_2$As$_2$ from
  Quantum Oscillations in Magnetostriction}},}\ }\href@noop {} {\bibfield
  {journal} {\bibinfo  {journal} {JPS Conf. Proc.}\ }\textbf {\bibinfo {volume}
  {3}},\ \bibinfo {pages} {015007} (\bibinfo {year} {2014})}\BibitemShut
  {NoStop}%
\bibitem [{\citenamefont {{Delong Fang, Xun Shi, Zengyi Du, Pierre Richard,
  Huan Yang, X. X. Wu, Peng Zhang, Tian Qian, Xiaxin Ding, Zhenyu Wang, T. K.
  Kim, M. Hoesch, Aifeng Wang, Xianhui Chen, Jiangping Hu, Hong Ding and Hai-Hu
  Wen}}(2015)}]{DL_Fang_vHs}%
  \BibitemOpen
  \bibfield  {author} {\bibinfo {author} {\bibnamefont {{Delong Fang, Xun Shi,
  Zengyi Du, Pierre Richard, Huan Yang, X. X. Wu, Peng Zhang, Tian Qian, Xiaxin
  Ding, Zhenyu Wang, T. K. Kim, M. Hoesch, Aifeng Wang, Xianhui Chen, Jiangping
  Hu, Hong Ding and Hai-Hu Wen}}},\ }\bibfield  {title} {\enquote {\bibinfo
  {title} {{Observation of a Van Hove singularity and implication for
  strong-coupling induced Cooper pairing in KFe$_2$As$_2$}},}\ }\href@noop {}
  {\bibfield  {journal} {\bibinfo  {journal} {Phys. Rev. B}\ }\textbf {\bibinfo
  {volume} {92}},\ \bibinfo {pages} {144513} (\bibinfo {year}
  {2015})}\BibitemShut {NoStop}%
\bibitem [{\citenamefont {{F. Hardy, A. E. B\"{o}hmer, L. de' Medici, M.
  Capone, G. Giovannetti, R. Eder, L. Wang, M. He, T. Wolf, P. Schweiss, R.
  Heid, A. Herbig, P. Adelmann, R. A. Fisher and C.
  Meingast}}(2016)}]{Hardy_PRB94}%
  \BibitemOpen
  \bibfield  {author} {\bibinfo {author} {\bibnamefont {{F. Hardy, A. E.
  B\"{o}hmer, L. de' Medici, M. Capone, G. Giovannetti, R. Eder, L. Wang, M.
  He, T. Wolf, P. Schweiss, R. Heid, A. Herbig, P. Adelmann, R. A. Fisher and
  C. Meingast}}},\ }\bibfield  {title} {\enquote {\bibinfo {title} {{Strong
  correlations, strong coupling, and $s$-wave superconductivity in hole-doped
  BaFe$_2$As$_2$ single crystals}},}\ }\href@noop {} {\bibfield  {journal}
  {\bibinfo  {journal} {Phys. Rev. B}\ }\textbf {\bibinfo {volume} {94}},\
  \bibinfo {pages} {205113} (\bibinfo {year} {2016})}\BibitemShut {NoStop}%
\bibitem [{\citenamefont {{Felix Eilers, Kai Grube, Diego A. Zocco, Thomas
  Wolf, Michael Merz, Peter Schweiss, Rolf Heid, Robert Eder, Rong Yu, Jian-Xin
  Zhu, Qimiao Si, Takasada Shibauchi and Hilbert v.
  L\"{o}hneysen}}(2016)}]{Eilers_PRL116}%
  \BibitemOpen
  \bibfield  {author} {\bibinfo {author} {\bibnamefont {{Felix Eilers, Kai
  Grube, Diego A. Zocco, Thomas Wolf, Michael Merz, Peter Schweiss, Rolf Heid,
  Robert Eder, Rong Yu, Jian-Xin Zhu, Qimiao Si, Takasada Shibauchi and Hilbert
  v. L\"{o}hneysen}}},\ }\bibfield  {title} {\enquote {\bibinfo {title}
  {{Strain-Driven Approach to Quantum Criticality in $A$Fe$_2$As$_2$ with $A$ =
  K, Rb, and Cs}},}\ }\href@noop {} {\bibfield  {journal} {\bibinfo  {journal}
  {Phys. Rev. Lett.}\ }\textbf {\bibinfo {volume} {116}},\ \bibinfo {pages}
  {237003} (\bibinfo {year} {2016})}\BibitemShut {NoStop}%
\bibitem [{\citenamefont {{T. Sato, K. Nakayama, Y. Sekiba, P. Richard, Y.-M.
  Xu, S. Souma, T. Takahashi, G. F. Chen, J. L. Luo, N. L. Wang and H.
  Ding}}(2009)}]{Sato_PRL2009}%
  \BibitemOpen
  \bibfield  {author} {\bibinfo {author} {\bibnamefont {{T. Sato, K. Nakayama,
  Y. Sekiba, P. Richard, Y.-M. Xu, S. Souma, T. Takahashi, G. F. Chen, J. L.
  Luo, N. L. Wang and H. Ding}}},\ }\bibfield  {title} {\enquote {\bibinfo
  {title} {{Band Structure and Fermi Surface of an Extremely Overdoped
  Iron-Based Superconductor KFe$_2$As$_2$}},}\ }\href@noop {} {\bibfield
  {journal} {\bibinfo  {journal} {Phys. Rev. Lett.}\ }\textbf {\bibinfo
  {volume} {103}},\ \bibinfo {pages} {047002} (\bibinfo {year}
  {2009})}\BibitemShut {NoStop}%
\bibitem [{\citenamefont {{T. Yoshida, I. Nishi, A. Fujimori, M. Yi, R. G.
  Moore, D.-H. Lu, Z.-X. Shen, K. Kihou, P. M. Shirage, H. Kito, C. H. Lee, A.
  Iyo, H. Eisaki and H. Harima}}(2011)}]{Yoshida_JCPS72}%
  \BibitemOpen
  \bibfield  {author} {\bibinfo {author} {\bibnamefont {{T. Yoshida, I. Nishi,
  A. Fujimori, M. Yi, R. G. Moore, D.-H. Lu, Z.-X. Shen, K. Kihou, P. M.
  Shirage, H. Kito, C. H. Lee, A. Iyo, H. Eisaki and H. Harima}}},\ }\bibfield
  {title} {\enquote {\bibinfo {title} {{Fermi surfaces and quasi-particle band
  dispersions of the iron pnictides superconductor KFe$_2$As$_2$ observed by
  angle-resolved photoemission spectroscopy}},}\ }\href@noop {} {\bibfield
  {journal} {\bibinfo  {journal} {J. Chem. Phys. Sol.}\ }\textbf {\bibinfo
  {volume} {72}},\ \bibinfo {pages} {465} (\bibinfo {year} {2011})}\BibitemShut
  {NoStop}%
\bibitem [{\citenamefont {{T. Yoshida, S.-I. Ideta, I. Nishi, A. Fujimori, M.
  Yi, R. G. Moore, S.-K. Mo, D. Lu, Z.-X. Shen, Z. Hussain, K. Kihou, P. M.
  Shirage, H. Kito, C.-H. Lee, A. Iyo, H. Eisaki and H.
  Harima}}(2014)}]{Yoshida_FP2}%
  \BibitemOpen
  \bibfield  {author} {\bibinfo {author} {\bibnamefont {{T. Yoshida, S.-I.
  Ideta, I. Nishi, A. Fujimori, M. Yi, R. G. Moore, S.-K. Mo, D. Lu, Z.-X.
  Shen, Z. Hussain, K. Kihou, P. M. Shirage, H. Kito, C.-H. Lee, A. Iyo, H.
  Eisaki and H. Harima}}},\ }\bibfield  {title} {\enquote {\bibinfo {title}
  {{Orbital character and electron correlation effects on two- and
  three-dimensional Fermi surfaces in KFe$_2$As$_2$ revealed by angle-resolved
  photoemission spectroscopy}},}\ }\href@noop {} {\bibfield  {journal}
  {\bibinfo  {journal} {Front. Phys.}\ }\textbf {\bibinfo {volume} {2}},\
  \bibinfo {pages} {17} (\bibinfo {year} {2014})}\BibitemShut {NoStop}%
\bibitem [{\citenamefont {{S. Kong, D. Y. Liu, S. T. Cui, S. L. Ju, A. F. Wang,
  X. G. Luo, L. J. Zou, X. H. Chen, G. B. Zhang and Z.
  Sun}}(2015)}]{Kong_PRB92}%
  \BibitemOpen
  \bibfield  {author} {\bibinfo {author} {\bibnamefont {{S. Kong, D. Y. Liu, S.
  T. Cui, S. L. Ju, A. F. Wang, X. G. Luo, L. J. Zou, X. H. Chen, G. B. Zhang
  and Z. Sun}}},\ }\bibfield  {title} {\enquote {\bibinfo {title} {{Electronic
  structure in a one-Fe Brillouin zone of the iron pnictide superconductors
  CsFe$_2$As$_2$ and RbFe$_2$As$_2$}},}\ }\href@noop {} {\bibfield  {journal}
  {\bibinfo  {journal} {Phys. Rev. B}\ }\textbf {\bibinfo {volume} {92}},\
  \bibinfo {pages} {184512} (\bibinfo {year} {2015})}\BibitemShut {NoStop}%
\bibitem [{\citenamefont {{T. Terashima, M. Kimata, N. Kurita, H. Satsukawa, A.
  Harada, K. Hazama, M. Imai, A. Sato, K. Kihou, C.-H. Lee, H. Kito, H. Eisaki,
  A. Iyo, T. Saito, H. Fukuzawa, Y. Kohori, H. Harima and S.
  Uji}}(2010)}]{Terashima_JPSJ79}%
  \BibitemOpen
  \bibfield  {author} {\bibinfo {author} {\bibnamefont {{T. Terashima, M.
  Kimata, N. Kurita, H. Satsukawa, A. Harada, K. Hazama, M. Imai, A. Sato, K.
  Kihou, C.-H. Lee, H. Kito, H. Eisaki, A. Iyo, T. Saito, H. Fukuzawa, Y.
  Kohori, H. Harima and S. Uji}}},\ }\bibfield  {title} {\enquote {\bibinfo
  {title} {{Fermi Surface and Mass Enhancement in KFe$_2$As$_2$ from de
  Haas-van Alphen Effect Measurements}},}\ }\href@noop {} {\bibfield  {journal}
  {\bibinfo  {journal} {J. Phys. Soc. Jpn}\ }\textbf {\bibinfo {volume} {79}},\
  \bibinfo {pages} {053702} (\bibinfo {year} {2010})}\BibitemShut {NoStop}%
\bibitem [{\citenamefont {{T. Terashima, N. Kurita, M. Kimata, M. Tomita, S.
  Tsuchiya, M. Imai, A. Sato, K. Kihou, C.-H. Lee, H. Kito, H. Eisaki, A. Iyo,
  T. Saito, H. Fukazawa, Y. Kohori, H. Harima and S.
  Uji}}(2013)}]{Terashima_PRB87}%
  \BibitemOpen
  \bibfield  {author} {\bibinfo {author} {\bibnamefont {{T. Terashima, N.
  Kurita, M. Kimata, M. Tomita, S. Tsuchiya, M. Imai, A. Sato, K. Kihou, C.-H.
  Lee, H. Kito, H. Eisaki, A. Iyo, T. Saito, H. Fukazawa, Y. Kohori, H. Harima
  and S. Uji}}},\ }\bibfield  {title} {\enquote {\bibinfo {title} {{Fermi
  surface in KFe$_2$As$_2$ determined via de Haas-van Alphen oscillation
  measurements}},}\ }\href@noop {} {\bibfield  {journal} {\bibinfo  {journal}
  {Phys. Rev. B}\ }\textbf {\bibinfo {volume} {87}},\ \bibinfo {pages} {224512}
  (\bibinfo {year} {2013})}\BibitemShut {NoStop}%
\bibitem [{\citenamefont {{T. Terashima, K. Kihou, K. Sugii, N. Kikugawa, T.
  Matsumoto, S. Ishida, C.-H. Lee, A. Iyo, H. Eisaki and S.
  Uji}}(2014)}]{Terashima_PRB89}%
  \BibitemOpen
  \bibfield  {author} {\bibinfo {author} {\bibnamefont {{T. Terashima, K.
  Kihou, K. Sugii, N. Kikugawa, T. Matsumoto, S. Ishida, C.-H. Lee, A. Iyo, H.
  Eisaki and S. Uji}}},\ }\bibfield  {title} {\enquote {\bibinfo {title} {{Two
  distinct superconducting states in KFe$_2$As$_2$ under high pressure}},}\
  }\href@noop {} {\bibfield  {journal} {\bibinfo  {journal} {Phys. Rev. B}\
  }\textbf {\bibinfo {volume} {89}},\ \bibinfo {pages} {134520} (\bibinfo
  {year} {2014})}\BibitemShut {NoStop}%
\bibitem [{Note1()}]{Note1}%
  \BibitemOpen
  \bibinfo {note} {We caution that the axes in Ref. \cite {Backes_NJP16} are
  rotated by 45$^{\circ }$ with respect to ours}\BibitemShut {NoStop}%
\bibitem [{\citenamefont {{Steffen Backes, Daniel Guterding, Harald O. Jeschke
  and Roser Valent\'{i}}}(2014)}]{Backes_NJP16}%
  \BibitemOpen
  \bibfield  {author} {\bibinfo {author} {\bibnamefont {{Steffen Backes, Daniel
  Guterding, Harald O. Jeschke and Roser Valent\'{i}}}},\ }\bibfield  {title}
  {\enquote {\bibinfo {title} {{Electronic structure and de Haas-van Alphen
  frequencies in KFe$_2$As$_2$ within LDA+DMFT}},}\ }\href@noop {} {\bibfield
  {journal} {\bibinfo  {journal} {N. J. Phys.}\ }\textbf {\bibinfo {volume}
  {16}},\ \bibinfo {pages} {083025} (\bibinfo {year} {2014})}\BibitemShut
  {NoStop}%
\bibitem [{\citenamefont {{P. Richard, T. Qian and H.
  Ding}}(2015)}]{RichardJPCM27review}%
  \BibitemOpen
  \bibfield  {author} {\bibinfo {author} {\bibnamefont {{P. Richard, T. Qian
  and H. Ding}}},\ }\bibfield  {title} {\enquote {\bibinfo {title} {{ARPES
  measurements of the superconducting gap of Fe-based superconductors and their
  implications to the pairing mechanism}},}\ }\href@noop {} {\bibfield
  {journal} {\bibinfo  {journal} {J. Phys.: Condens. Matter}\ }\textbf
  {\bibinfo {volume} {27}},\ \bibinfo {pages} {293203} (\bibinfo {year}
  {2015})}\BibitemShut {NoStop}%
\bibitem [{\citenamefont {{P. Richard, A. van Roekeghem, B. Q. Lv, Tian Qian,
  T. K. Kim, M. Hoesch, J.-P. Hu, Athena S. Sefat, Silke Biermann and H.
  Ding}}(2017)}]{RichardPRB95}%
  \BibitemOpen
  \bibfield  {author} {\bibinfo {author} {\bibnamefont {{P. Richard, A. van
  Roekeghem, B. Q. Lv, Tian Qian, T. K. Kim, M. Hoesch, J.-P. Hu, Athena S.
  Sefat, Silke Biermann and H. Ding}}},\ }\bibfield  {title} {\enquote
  {\bibinfo {title} {{Is BaCr$_2$As$_2$ symmetrical to BaFe$_2$As$_2$ with
  respect to half 3d shell filling?}}}\ }\href@noop {} {\bibfield  {journal}
  {\bibinfo  {journal} {Phys. Rev. B}\ }\textbf {\bibinfo {volume} {95}},\
  \bibinfo {pages} {184516} (\bibinfo {year} {2017})}\BibitemShut {NoStop}%
\bibitem [{\citenamefont {{Jayita Nayak, Kai Filsinger, Gerhard H. Fecher,
  Stanislav Chadov, J\'{a}n Min\'{a}r, Emile D. L. Rienks, Bernd B\"{u}chner,
  Stuart P. Parkine, J\"{o}rg Finka and Claudia
  Felser}}(2017)}]{Nayak_PNAS114}%
  \BibitemOpen
  \bibfield  {author} {\bibinfo {author} {\bibnamefont {{Jayita Nayak, Kai
  Filsinger, Gerhard H. Fecher, Stanislav Chadov, J\'{a}n Min\'{a}r, Emile D.
  L. Rienks, Bernd B\"{u}chner, Stuart P. Parkine, J\"{o}rg Finka and Claudia
  Felser}}},\ }\bibfield  {title} {\enquote {\bibinfo {title} {{Observation of
  a remarkable reduction of correlation effects in BaCr$_2$As$_2$ by ARPES}},}\
  }\href@noop {} {\bibfield  {journal} {\bibinfo  {journal} {Proc. Natl. Acad.
  Sci. USA}\ }\textbf {\bibinfo {volume} {114}},\ \bibinfo {pages} {12425}
  (\bibinfo {year} {2017})}\BibitemShut {NoStop}%
\bibitem [{\citenamefont {{J. Hu, C. Le and X. Wu}}(2015)}]{JP_HuPRX5}%
  \BibitemOpen
  \bibfield  {author} {\bibinfo {author} {\bibnamefont {{J. Hu, C. Le and X.
  Wu}}},\ }\bibfield  {title} {\enquote {\bibinfo {title} {{Predicting
  Unconventional High-Temperature Superconductors in Trigonal Bipyramidal
  Coordinations}},}\ }\href@noop {} {\bibfield  {journal} {\bibinfo  {journal}
  {Phys. Rev. X}\ }\textbf {\bibinfo {volume} {101}},\ \bibinfo {pages}
  {041012} (\bibinfo {year} {2015})}\BibitemShut {NoStop}%
\bibitem [{\citenamefont {{E. Razzoli, C. E. Matt, M. Kobayashi, X.-P. Wang, V.
  N. Strocov, A. van Roekeghem, S. Biermann, N. C. Plumb, M. Radovic, T.
  Schmitt, C. Capan, Z. Fisk, P. Richard, H. Ding, P. Aebi, J. Mesot and M.
  Shi}}(2015)}]{Razzoli_PRB91}%
  \BibitemOpen
  \bibfield  {author} {\bibinfo {author} {\bibnamefont {{E. Razzoli, C. E.
  Matt, M. Kobayashi, X.-P. Wang, V. N. Strocov, A. van Roekeghem, S. Biermann,
  N. C. Plumb, M. Radovic, T. Schmitt, C. Capan, Z. Fisk, P. Richard, H. Ding,
  P. Aebi, J. Mesot and M. Shi}}},\ }\bibfield  {title} {\enquote {\bibinfo
  {title} {{Tuning electronic correlations in transition metal pnictides:
  Chemistry beyond the valence count}},}\ }\href@noop {} {\bibfield  {journal}
  {\bibinfo  {journal} {Phys. Rev. B}\ }\textbf {\bibinfo {volume} {91}},\
  \bibinfo {pages} {214502} (\bibinfo {year} {2015})}\BibitemShut {NoStop}%
\bibitem [{\citenamefont {{Steffen Backes, Harald O. Jeschke and Roser
  Valent\'{i}}}(2015)}]{Backes_PRB92}%
  \BibitemOpen
  \bibfield  {author} {\bibinfo {author} {\bibnamefont {{Steffen Backes, Harald
  O. Jeschke and Roser Valent\'{i}}}},\ }\bibfield  {title} {\enquote {\bibinfo
  {title} {{Microscopic nature of correlations in multiorbital $A$Fe$_2$As$_2$
  ($A$ = K, Rb, Cs): Hund's coupling versus Coulomb repulsion}},}\ }\href@noop
  {} {\bibfield  {journal} {\bibinfo  {journal} {Phys. Rev. B}\ }\textbf
  {\bibinfo {volume} {92}},\ \bibinfo {pages} {195128} (\bibinfo {year}
  {2015})}\BibitemShut {NoStop}%
\bibitem [{\citenamefont {{Stefan-Ludwig Drechsler, Steve Johnston, Vadim
  Grinenko, Jan M. Tomczak and Helge Rosner}}(2017)}]{Drechsler_PSSB254}%
  \BibitemOpen
  \bibfield  {author} {\bibinfo {author} {\bibnamefont {{Stefan-Ludwig
  Drechsler, Steve Johnston, Vadim Grinenko, Jan M. Tomczak and Helge
  Rosner}}},\ }\bibfield  {title} {\enquote {\bibinfo {title} {{Constraints on
  the total coupling strength to bosons in the iron based superconductors}},}\
  }\href@noop {} {\bibfield  {journal} {\bibinfo  {journal} {Phys. Status Sol.
  B}\ }\textbf {\bibinfo {volume} {254}},\ \bibinfo {pages} {1700006} (\bibinfo
  {year} {2017})}\BibitemShut {NoStop}%
\bibitem [{\citenamefont {{S.-L. Drechsler, H. Rosner, V. Grinenko, S.
  Aswartham, I. Morozov1, M. Liu, A. Boltalin, K. Kihou, C. H. Lee, T. Kim, D.
  Evtushinsky, J. M. Tomczak, S. Johnston and S.
  Borisenko}}(2018)}]{Drechsler_JSNM31}%
  \BibitemOpen
  \bibfield  {author} {\bibinfo {author} {\bibnamefont {{S.-L. Drechsler, H.
  Rosner, V. Grinenko, S. Aswartham, I. Morozov1, M. Liu, A. Boltalin, K.
  Kihou, C. H. Lee, T. Kim, D. Evtushinsky, J. M. Tomczak, S. Johnston and S.
  Borisenko}}},\ }\bibfield  {title} {\enquote {\bibinfo {title} {{Mass
  Enhancements and Band Shifts in Strongly Hole-Overdoped Fe-Based Pnictide
  Superconductors: KFe$_2$As$_2$ and CsFe$_2$As$_2$}},}\ }\href@noop {}
  {\bibfield  {journal} {\bibinfo  {journal} {J. Supercond. Novel Magn.}\
  }\textbf {\bibinfo {volume} {31}},\ \bibinfo {pages} {777} (\bibinfo {year}
  {2018})}\BibitemShut {NoStop}%
\bibitem [{\citenamefont {{N. Xu, C. E. Matt, P. Richard, A. van Roekeghem, S.
  Biermann, X. Shi, S.-F. Wu, H. W. Liu, D. Chen, T. Qian, N. C. Plumb, M.
  Radovi\'{c}, Hangdong Wang, Qianhui Mao, Jianhua Du, Minghu Fang, J. Mesot,
  H. Ding and M. Shi}}(2015)}]{Nan_XuPRB92}%
  \BibitemOpen
  \bibfield  {author} {\bibinfo {author} {\bibnamefont {{N. Xu, C. E. Matt, P.
  Richard, A. van Roekeghem, S. Biermann, X. Shi, S.-F. Wu, H. W. Liu, D. Chen,
  T. Qian, N. C. Plumb, M. Radovi\'{c}, Hangdong Wang, Qianhui Mao, Jianhua Du,
  Minghu Fang, J. Mesot, H. Ding and M. Shi}}},\ }\bibfield  {title} {\enquote
  {\bibinfo {title} {{Camelback-shaped band reconciles heavy-electron behavior
  with weak electronic Coulomb correlations in superconducting
  TlNi$_2$Se$_2$}},}\ }\href@noop {} {\bibfield  {journal} {\bibinfo  {journal}
  {Phys. Rev. B}\ }\textbf {\bibinfo {volume} {92}},\ \bibinfo {pages}
  {081116(R)} (\bibinfo {year} {2015})}\BibitemShut {NoStop}%
\bibitem [{\citenamefont {{Hangdong Wang, Chiheng Dong, Qianhui Mao, Rajwali
  Khan, Xi Zhou, Chenxia Li, Bin Chen, Jinhu Yang, Qiping Su and Minghu
  Fang}}(2013)}]{H_WangPRL111}%
  \BibitemOpen
  \bibfield  {author} {\bibinfo {author} {\bibnamefont {{Hangdong Wang, Chiheng
  Dong, Qianhui Mao, Rajwali Khan, Xi Zhou, Chenxia Li, Bin Chen, Jinhu Yang,
  Qiping Su and Minghu Fang}}},\ }\bibfield  {title} {\enquote {\bibinfo
  {title} {{Multiband Superconductivity of Heavy Electrons in a TlNi$_2$Se$_2$
  Single Crystal}},}\ }\href@noop {} {\bibfield  {journal} {\bibinfo  {journal}
  {Phys. Rev. Lett.}\ }\textbf {\bibinfo {volume} {111}},\ \bibinfo {pages}
  {207001} (\bibinfo {year} {2013})}\BibitemShut {NoStop}%
\bibitem [{\citenamefont {{N. Xu, P. Richard, X. Shi, A. van Roekeghem, T.
  Qian, E. Razzoli, E. Rienks, G.-F. Chen, E. Ieki, K. Nakayama, T. Sato, T.
  Takahashi, M. Shi and H. Ding}}(2013)}]{Nan_XuPRB88}%
  \BibitemOpen
  \bibfield  {author} {\bibinfo {author} {\bibnamefont {{N. Xu, P. Richard, X.
  Shi, A. van Roekeghem, T. Qian, E. Razzoli, E. Rienks, G.-F. Chen, E. Ieki,
  K. Nakayama, T. Sato, T. Takahashi, M. Shi and H. Ding}}},\ }\bibfield
  {title} {\enquote {\bibinfo {title} {{Possible nodal superconducting gap and
  Lifshitz transition in heavily hole-doped
  Ba$_{0.1}$K$_{0.9}$Fe$_2$As$_2$}},}\ }\href@noop {} {\bibfield  {journal}
  {\bibinfo  {journal} {Phys. Rev. B}\ }\textbf {\bibinfo {volume} {88}},\
  \bibinfo {pages} {220508(R)} (\bibinfo {year} {2013})}\BibitemShut {NoStop}%
\bibitem [{Note2()}]{Note2}%
  \BibitemOpen
  \bibinfo {note} {See e.g. Fig.~3 of Ref.~\cite {TaufourPRB89}, where
  deviations from $T^2$ behavior of the resistivity appear at temperatures of
  the order of 60-70 K.}\BibitemShut {Stop}%
\bibitem [{\citenamefont {{L.-K. Zeng, P. Richard, A. van Roekeghem, J.-X. Yin,
  S.-F. Wu, Z. G. Chen, N. L. Wang, S. Biermann, T. Qian and H.
  Ding}}(2016)}]{LK_ZengPRB94}%
  \BibitemOpen
  \bibfield  {author} {\bibinfo {author} {\bibnamefont {{L.-K. Zeng, P.
  Richard, A. van Roekeghem, J.-X. Yin, S.-F. Wu, Z. G. Chen, N. L. Wang, S.
  Biermann, T. Qian and H. Ding}}},\ }\bibfield  {title} {\enquote {\bibinfo
  {title} {{Angle-resolved spectroscopy study of Ni-based superconductor
  SrNi$_2$As$_2$}},}\ }\href@noop {} {\bibfield  {journal} {\bibinfo  {journal}
  {Phys. Rev. B}\ }\textbf {\bibinfo {volume} {94}},\ \bibinfo {pages} {024524}
  (\bibinfo {year} {2016})}\BibitemShut {NoStop}%
\bibitem [{\citenamefont {{N. Xu, P. Richard, A. van Roekeghem, P. Zhang, H.
  Miao, W.-L. Zhang, T. Qian, M. Ferrero, A. S. Sefat, S. Biermann and H.
  Ding}}(2013)}]{Nan_XuPRX3}%
  \BibitemOpen
  \bibfield  {author} {\bibinfo {author} {\bibnamefont {{N. Xu, P. Richard, A.
  van Roekeghem, P. Zhang, H. Miao, W.-L. Zhang, T. Qian, M. Ferrero, A. S.
  Sefat, S. Biermann and H. Ding}}},\ }\href@noop {} {\bibfield  {journal}
  {\bibinfo  {journal} {Phys. Rev. X}\ }\textbf {\bibinfo {volume} {3}},\
  \bibinfo {pages} {011006} (\bibinfo {year} {2013})}\BibitemShut {NoStop}%
\bibitem [{\citenamefont {{V. Brouet, Ping-Hui Lin, Y. Texier, J. Bobroff, A.
  Taleb-Ibrahimi, P. Le F\`{e}vre, F. Bertran, M. Casula, P. Werner, S.
  Biermann, F. Rullier-Albenque, A. Forget, and D.
  Colson}}(2013)}]{Brouet_PRL110}%
  \BibitemOpen
  \bibfield  {author} {\bibinfo {author} {\bibnamefont {{V. Brouet, Ping-Hui
  Lin, Y. Texier, J. Bobroff, A. Taleb-Ibrahimi, P. Le F\`{e}vre, F. Bertran,
  M. Casula, P. Werner, S. Biermann, F. Rullier-Albenque, A. Forget, and D.
  Colson}}},\ }\href@noop {} {\bibfield  {journal} {\bibinfo  {journal} {Phys.
  Rev. Lett.}\ }\textbf {\bibinfo {volume} {110}},\ \bibinfo {pages} {167002}
  (\bibinfo {year} {2013})}\BibitemShut {NoStop}%
\bibitem [{\citenamefont {{R. S. Dhaka, S. E. Hahn, E. Razzoli, Rui Jiang, M.
  Shi, B. N. Harmon, A. Thaler, S. L. Bud'ko, P. C. Canfield and Adam
  Kaminski}}(2013)}]{Dhaka_PRL110}%
  \BibitemOpen
  \bibfield  {author} {\bibinfo {author} {\bibnamefont {{R. S. Dhaka, S. E.
  Hahn, E. Razzoli, Rui Jiang, M. Shi, B. N. Harmon, A. Thaler, S. L. Bud'ko,
  P. C. Canfield and Adam Kaminski}}},\ }\href@noop {} {\bibfield  {journal}
  {\bibinfo  {journal} {Phys. Rev. Lett.}\ }\textbf {\bibinfo {volume} {110}},\
  \bibinfo {pages} {067002} (\bibinfo {year} {2013})}\BibitemShut {NoStop}%
\bibitem [{\citenamefont {{Ping-Hui Lin, Y. Texier, A. Taleb-Ibrahimi, P. Le
  F\`{e}vre, F. Bertran, E. Giannini, M. Grioni and V.
  Brouet}}(2013)}]{PH_LinPRL111}%
  \BibitemOpen
  \bibfield  {author} {\bibinfo {author} {\bibnamefont {{Ping-Hui Lin, Y.
  Texier, A. Taleb-Ibrahimi, P. Le F\`{e}vre, F. Bertran, E. Giannini, M.
  Grioni and V. Brouet}}},\ }\href@noop {} {\bibfield  {journal} {\bibinfo
  {journal} {Phys. Rev. Lett.}\ }\textbf {\bibinfo {volume} {111}},\ \bibinfo
  {pages} {217002} (\bibinfo {year} {2013})}\BibitemShut {NoStop}%
\bibitem [{\citenamefont {{I. M. Lifshitz}}(1960)}]{Lifshitz_JETP11}%
  \BibitemOpen
  \bibfield  {author} {\bibinfo {author} {\bibnamefont {{I. M. Lifshitz}}},\
  }\href@noop {} {\bibfield  {journal} {\bibinfo  {journal} {Sov. Phys. JETP}\
  }\textbf {\bibinfo {volume} {11}},\ \bibinfo {pages} {1130} (\bibinfo {year}
  {1960})}\BibitemShut {NoStop}%
\bibitem [{\citenamefont {{J. Fink}}(2016)}]{FinkEPL113}%
  \BibitemOpen
  \bibfield  {author} {\bibinfo {author} {\bibnamefont {{J. Fink}}},\
  }\bibfield  {title} {\enquote {\bibinfo {title} {{Influence of Lifshitz
  transitions and correlation effects on the scattering rates of the charge
  carriers in iron-based superconductors}},}\ }\href@noop {} {\bibfield
  {journal} {\bibinfo  {journal} {Europhys. Lett.}\ }\textbf {\bibinfo {volume}
  {113}},\ \bibinfo {pages} {27002} (\bibinfo {year} {2016})}\BibitemShut
  {NoStop}%
\bibitem [{\citenamefont {{R. Nourafkan and S.
  Acheche}}(2018)}]{Nourafkan_Knight_shift}%
  \BibitemOpen
  \bibfield  {author} {\bibinfo {author} {\bibnamefont {{R. Nourafkan and S.
  Acheche}}},\ }\bibfield  {title} {\enquote {\bibinfo {title} {{Temperature
  dependence of NMR Knight shift in pnictides: proximity to a van Hove
  singularity}},}\ }\href@noop {} {\bibfield  {journal} {\bibinfo  {journal}
  {arXiv:1807.09683v1}\ } (\bibinfo {year} {2018})}\BibitemShut {NoStop}%
\bibitem [{\citenamefont {{Ambroise van Roekeghem, Jes\'{u}s Carrete and
  Natalio Mingo}}(2016)}]{vanRoekeghem_PRB94}%
  \BibitemOpen
  \bibfield  {author} {\bibinfo {author} {\bibnamefont {{Ambroise van
  Roekeghem, Jes\'{u}s Carrete and Natalio Mingo}}},\ }\bibfield  {title}
  {\enquote {\bibinfo {title} {{Anomalous thermal conductivity and suppression
  of negative thermal expansion in ScF$_3$}},}\ }\href@noop {} {\bibfield
  {journal} {\bibinfo  {journal} {Phys. Rev. B}\ }\textbf {\bibinfo {volume}
  {94}},\ \bibinfo {pages} {020303(R)} (\bibinfo {year} {2016})}\BibitemShut
  {NoStop}%
\bibitem [{\citenamefont {{P. B. Allen}}(1978)}]{Allan_PRB17}%
  \BibitemOpen
  \bibfield  {author} {\bibinfo {author} {\bibnamefont {{P. B. Allen}}},\
  }\bibfield  {title} {\enquote {\bibinfo {title} {{New method for solving
  Boltzmann's equation for electrons in metals}},}\ }\href@noop {} {\bibfield
  {journal} {\bibinfo  {journal} {Phys. Rev. B}\ }\textbf {\bibinfo {volume}
  {17}},\ \bibinfo {pages} {3725} (\bibinfo {year} {1978})}\BibitemShut
  {NoStop}%
\bibitem [{\citenamefont {{R. Mittal, R. Heid, A. Bosak, T. R. Forrest, S. L.
  Chaplot, D. Lamago, D. Reznik, K.-P. Bohnen, Y. Su, N. Kumar, S. K. Dhar, A.
  Thamizhavel, Ch. R\"{u}egg, M. Krisch, D. F. McMorrow, Th. Brueckel and L.
  Pintschovius}}(2010)}]{Mittal_PRB81}%
  \BibitemOpen
  \bibfield  {author} {\bibinfo {author} {\bibnamefont {{R. Mittal, R. Heid, A.
  Bosak, T. R. Forrest, S. L. Chaplot, D. Lamago, D. Reznik, K.-P. Bohnen, Y.
  Su, N. Kumar, S. K. Dhar, A. Thamizhavel, Ch. R\"{u}egg, M. Krisch, D. F.
  McMorrow, Th. Brueckel and L. Pintschovius}}},\ }\bibfield  {title} {\enquote
  {\bibinfo {title} {{Pressure dependence of phonon modes across the tetragonal
  to collapsed-tetragonal phase transition in CaFe$_2$As$_2$}},}\ }\href@noop
  {} {\bibfield  {journal} {\bibinfo  {journal} {Phys. Rev. B}\ }\textbf
  {\bibinfo {volume} {81}},\ \bibinfo {pages} {144502} (\bibinfo {year}
  {2010})}\BibitemShut {NoStop}%
\bibitem [{\citenamefont {{Ambroise van Roekeghem, Pierre Richard, Xun Shi,
  Shangfei Wu, Lingkun Zeng, Bayrammurad Saparov, Yoshiyuki Ohtsubo, Tian Qian,
  Athena S. Sefat, Silke Biermann and Hong Ding}}(2016)}]{vanRoekeghem_PRB93}%
  \BibitemOpen
  \bibfield  {author} {\bibinfo {author} {\bibnamefont {{Ambroise van
  Roekeghem, Pierre Richard, Xun Shi, Shangfei Wu, Lingkun Zeng, Bayrammurad
  Saparov, Yoshiyuki Ohtsubo, Tian Qian, Athena S. Sefat, Silke Biermann and
  Hong Ding}}},\ }\bibfield  {title} {\enquote {\bibinfo {title} {{Tetragonal
  and collapsed-tetragonal phases of CaFe$_2$As$_2$: A view from angle-resolved
  photoemission and dynamical mean-field theory}},}\ }\href@noop {} {\bibfield
  {journal} {\bibinfo  {journal} {Phys. Rev. B}\ }\textbf {\bibinfo {volume}
  {93}},\ \bibinfo {pages} {245139} (\bibinfo {year} {2016})}\BibitemShut
  {NoStop}%
\bibitem [{\citenamefont {{Valentin Taufour, Neda Foroozani, Makariy A.
  Tanatar, Jinhyuk Lim, Udhara Kaluarachchi, Stella K. Kim, Yong Liu, Thomas A.
  Lograsso, Vladimir G. Kogan, Ruslan Prozorov, Sergey L. Bud'ko, James S.
  Schilling, and Paul C. Canfield}}(2014)}]{TaufourPRB89}%
  \BibitemOpen
  \bibfield  {author} {\bibinfo {author} {\bibnamefont {{Valentin Taufour, Neda
  Foroozani, Makariy A. Tanatar, Jinhyuk Lim, Udhara Kaluarachchi, Stella K.
  Kim, Yong Liu, Thomas A. Lograsso, Vladimir G. Kogan, Ruslan Prozorov, Sergey
  L. Bud'ko, James S. Schilling, and Paul C. Canfield}}},\ }\bibfield  {title}
  {\enquote {\bibinfo {title} {{Upper critical field of KFe$_2$As$_2$ under
  pressure: A test for the change in the superconducting gap structure}},}\
  }\href@noop {} {\bibfield  {journal} {\bibinfo  {journal} {Phys. Rev. B}\
  }\textbf {\bibinfo {volume} {89}},\ \bibinfo {pages} {220509(R)} (\bibinfo
  {year} {2014})}\BibitemShut {NoStop}%
\end{thebibliography}%

\end{document}